\documentclass[letterpaper, 10 pt, conference]{ieeeconf}  % Comment this line out if you need a4paper

\IEEEoverridecommandlockouts                              % This command is only needed if 
\newif\ifarxiv
\arxivtrue

\usepackage{amsmath} % assumes amsmath package installed
\usepackage{amssymb}  % assumes amsmath package installed
\usepackage{optidef}
\usepackage{mathtools}
\usepackage{dsfont} 
\usepackage{comment}
\usepackage{xcolor}
\usepackage{balance}

\usepackage{amsthm}

\usepackage{algorithm}
\usepackage{algpseudocode}
\let\labelindent\relax
\usepackage{enumitem}
\usepackage{makecell}
\usepackage{url}
\usepackage{cite}
\usepackage{empheq}
\usepackage[colorinlistoftodos]{todonotes}
\usepackage[hidelinks]{hyperref}
\usepackage{tikz}
\usetikzlibrary{positioning,fit,arrows.meta,backgrounds,calc}

\theoremstyle{definition}
\newtheorem{definition}{Definition}

\newtheorem{theorem}{Theorem}
\newtheorem{lemma}{Lemma}
\theoremstyle{remark}
\newtheorem{remark}{Remark}

\newcommand{\rmdiag}[0]{\mathrm{diag}}
\newcommand{\rmpu}[0]{\mathrm{pu}}
\newcommand{\rmdo}[0]{\mathrm{do}}
\newcommand{\rmeq}[0]{\mathrm{eq}}
\newcommand{\rmiq}[0]{\mathrm{iq}}

\newcommand{\mathbbm}[1]{\mathds{#1}}

\newcommand{\AlgRef}[2]{\text{Alg.~\ref{#1}}\,(#2)}

\NewDocumentCommand{\SolRef}{m o o}{%
  \textnormal{solution of~\eqref{#1}}%
  \IfValueT{#2}{\IfBlankF{#2}{\textnormal{ at }#2}}%
  \IfValueT{#3}{\IfBlankF{#3}{\ \textnormal{(#3)}}}%
}

\makeatletter
\newcounter{algstep}[algorithm]
\renewcommand{\thealgstep}{\arabic{algstep}}
\newcommand{\AlgStep}[2]{%
  \Statex\refstepcounter{algstep}\label{#2}%
  \textbf{Step~\thealgstep:} #1%
  \def\@currentlabel{\theALG@line}}% restore line labels
\makeatother
\algrenewcommand{\algorithmiccomment}[1]{\hfill{\color{gray}\footnotesize// #1}}

\usepackage[colorinlistoftodos]{todonotes}

\title{Incentive-based Fleet Rebalancing in Vehicle-Sharing Systems via Equilibrium-Constrained Optimal Control}

\author{Daniel Tihanyi and Giuseppe Belgioioso% <-this % stops a space
	\thanks{The authors are with the Department of Decision and Control Systems, KTH Royal Institute of Technology, Stockholm, Sweden (emails: \texttt{\{tihanyi, giubel\}@kth.se}). This work was partially supported by the Wallenberg AI, Autonomous Systems and Software Program (WASP) funded by the Knut and Alice Wallenberg Foundation.} %
}

\begin{document}

\maketitle
\thispagestyle{empty}
\pagestyle{empty}

\begin{abstract}
We propose an incentive-based fleet rebalancing scheme for vehicle-sharing systems that uses fare discounts to steer users toward alternative pick-up locations. The discounts solve an optimal control problem that trades off users leaving the service against incentive costs over a day of operation. Users' responses to incentives are modeled as a Nash equilibrium of logit models coupled through shared vehicle-availability constraints, which makes the control problem a mathematical program with nested equilibrium constraints. We develop a first-order algorithm that backpropagates through the equilibria via nonsmooth implicit differentiation and enjoys convergence guarantees. In a case study that uses Helsinki bike-sharing data, our scheme reduces the number of leaving users by up to 99\% relative to no incentives, and outperforms heuristic policies while using up to 52\% less incentive.
\end{abstract}

%%%%%%%%%%%%%%%%%%%%%%%%%%%%%%%%%%%%%%%%%%%%%%%%%%%%%%%%%%%%%%%%%%%%%%%%%%%%%%%%
% INTRODUCTION  --  max k pages (now, n)
%%%%%%%%%%%%%%%%%%%%%%%%%%%%%%%%%%%%%%%%%%%%%%%%%%%%%%%%%%%%%%%%%%%%%%%%%%%%%%%%
\section{Introduction}

{% ===== Move i: Establishing the territory =====
Vehicle-sharing services, such as car- and bike-sharing, are increasingly promoted as a sustainable alternative to private vehicles in cities~\cite{Jorge2013-vs}. In new-generation one-way, instant-access, and free-floating systems, users pick up vehicles without reservation and drop them off anywhere in the operational area~\cite{Cepolina2014-tm, Jorge2013-vs}. However, since trip demand is asymmetric in space and time, vehicles accumulate in some areas and become scarce in others~\cite{Curtale2021-qm, Jorge2013-vs}. As users are reluctant to walk far to reach a vehicle~\cite{Curtale2021-qm, Herrmann2014-it}, such shortages leave demand unserved, pushing users to other transportation modes and suppressing future demand~\cite{Jorge2013-vs}. Keeping the fleet balanced is therefore central to their success.

% ===== Move ii: Establishing the niche =====
% --- Literature: rebalancing approaches ---
Fleet rebalancing is either operator-based or user-based~\cite{Cepolina2014-tm}. In operator-based relocation, the operator repositions vehicles directly, via staff or, in autonomous fleets, by commanding the vehicles, with decisions computed, e.g., via mixed-integer programming~\cite{Boyaci2015-ey}, fluid models~\cite{Hosseini2024-ps}, or approximate dynamic programming~\cite{Huang2024-at}. While effective, operator-based relocation is costly and limited by the available staff~\cite{Curtale2021-qm, Wang2021-db}. In user-based relocation, the operator instead offers incentives, such as fare discounts, steering users toward pick-up or drop-off locations that rebalance the fleet \cite{Curtale2021-qm, Wang2021-db}. Surveys show that many users accept moderate detours for discounts~\cite{Herrmann2014-it, Curtale2021-qm}, and incentives can substantially reduce the staff needed for rebalancing~\cite{Stokkink2021-ck}.

% --- Literature: incentive design and user modeling ---
Designing incentives requires predicting users' responses, for which logit models~\cite{Ben-Akiva1985-kw} are the standard tool in vehicle sharing~\cite{Liu2022-vm, Wang2021-db, Curtale2021-qm, Stokkink2021-ck}. Incentives have been designed per user request, trading off incentive costs against demand loss predicted via Markov chains~\cite{Stokkink2021-ck}, or via approximate dynamic programming policies that learn user preferences online~\cite{Liu2022-vm}. Other works take a system-level view, either jointly optimizing alternative destinations and fare discounts for all vehicles in use~\cite{Di-Febbraro2019-xd}, or setting incentives per station and time of day within a time-space network of the fleet dynamics~\cite{Wang2021-db}. However, these approaches predict each user's choice independently of the others~\cite{Di-Febbraro2019-xd}. In~\cite{Wang2021-db}, demand exceeding the available vehicles is rationed only afterward, so users who find no vehicle are lost rather than redirected, and incentives are optimized period by period via simulated annealing. In practice, users in different areas compete for the same vehicles, and neglecting this may yield incentives that attract more users to an area than it has vehicles, wasting budget~\cite{Wang2021-db}. Such competition among users has been captured via dynamic user equilibrium models~\cite{Wang2023-cm}, but only to evaluate predefined incentive strategies.

% --- Methodological challenge ---
Optimizing incentives against such equilibrium responses results instead in a mathematical program with equilibrium constraints (MPEC)~\cite{Luo1996-eh}. Over a day of operation, the equilibria are nested through the fleet dynamics, yielding an optimal control problem with equilibrium constraints (OCPEC) ~\cite{Lin2025-nip}. 
Numerical methods for OCPECs mostly replace each equilibrium with its complementarity conditions and solve the resulting problem as a single MPEC, via relaxation~\cite{scholtes2001convergence} or Newton-type methods~\cite{Lin2025-nip}. When each equilibrium is a single-valued Lipschitz function of its parameters, one can instead compute the chain forward and differentiate it backward in time, similar to closed-loop optimization of MPC~\cite{Zuliani2023-po}. For games, however, differentiating through equilibria has been limited to static interventions~\cite{Li2020-eg, Grontas2024-bh}.

% ===== Move iii: Occupying the niche =====
% --- Gap statement ---
In short, incentive schemes for vehicle sharing either neglect the competition among users or do not optimize over it, and, to the best of our knowledge, no first-order method differentiates through a chain of equilibria among competing users to optimize a finite-horizon cost with convergence guarantees. We close this gap via the second, backpropagation-based route. Our contributions are threefold.
\begin{itemize}
    \item We model the users' response to incentives as a generalized game in which the constrained logit models~\cite{Atarashi2025-wp} of users located in different regions are coupled by the shared availability of vehicles, capturing the competition for vehicles that existing incentive schemes neglect.
    
    \item We cast incentive-based rebalancing as an OCPEC, in which the incentives drive the fleet through the users' equilibrium response, and develop a first-order algorithm that backpropagates through the chain of equilibria to evaluate gradient information. Our approach builds on the algorithm in~\cite{Zuliani2023-po} and extends it to game-theoretic settings with entropic lower-level costs.
    
    \item We validate the proposed OCPEC approach in a case study on Helsinki city-bike data~\cite{hsl2025od}, comparing the optimized incentives against heuristic and myopic policies, in terms of users leaving the service and incentive usage. Code and data are publicly available\footnote{\url{https://github.com/tihanyidani/vehicle-sharing}}.

\end{itemize}

\subsection{Notation and preliminaries}
We denote by $\mathbb{N}$ the nonnegative integers and by $\mathbb{R}_{+}$ ($\mathbb{R}_{>0}$) the nonnegative (positive) reals. $\mathbbm{1}_N$ and $0_N$ are the all-ones and all-zeros vectors in $\mathbb{R}^N$, $0_{N \times M}$ the zero matrix, $I_N$ the identity, and $e_t \in \mathbb{R}^T$ the $t$-th canonical basis vector; dimensions are omitted when clear. Subscripts denote time, superscripts indices, and superscripts in parentheses iterations. $(v_1, \dots, v_n)$ and $(v_a)_{a \in \mathcal{A}}$ stack vectors or matrices vertically, the latter following the order of a finite ordered set $\mathcal{A}$. $\mathrm{diag}(v)$ is the diagonal matrix of $v$ and $\otimes$ the Kronecker product; vector inequalities, $\log(\cdot)$, $\exp(\cdot)$, and $\max(0, \cdot)$ act entry-wise. $\mathbb{S} \coloneq \{ p \in \mathbb{R}^{n}_{+} \mid \mathbbm{1}^\top p = 1 \}$ is the probability simplex and $\mathbb{S}^N$ its $N$-fold product. For a nonempty closed convex set $\mathcal{K} \subseteq \mathbb{R}^n$, $\Pi_{\mathcal{K}}(v) \coloneq \arg\min_{y \in \mathcal{K}} \|y - v\|_2$ is the projection onto $\mathcal{K}$ and $\mathcal{N}_{\mathcal{K}}(v) \coloneq \{ z \mid z^\top (y - v) \le 0,\ \forall y \in \mathcal{K} \}$ its normal cone at $v \in \mathcal{K}$. To differentiate nonsmooth maps, we use the following generalization of the Jacobian, where $\rightrightarrows$ denotes set-valued maps.

\begin{definition}[Conservative Jacobian {\cite{Bolte2021-ml}}]\label{def:conservative_jacobian}
    Given a locally Lipschitz function $f: \mathbb{R}^n \rightarrow \mathbb{R}^m$, a set-valued map $\mathcal{J}_x f: \mathbb{R}^n \rightrightarrows \mathbb{R}^{m \times n}$ with nonempty values, locally bounded and with closed graph, is a conservative Jacobian for $f$ if, for any absolutely continuous curve $\varphi: [0,1] \rightarrow \mathbb{R}^n$,
    \begin{equation}
        \frac{d}{dt} f(\varphi(t)) = J\, \dot{\varphi}(t), \quad \forall J \in \mathcal{J}_x f(\varphi(t)),
    \end{equation}
    for almost every $t \in [0,1]$. A locally Lipschitz function that admits a conservative Jacobian is called path-differentiable.
\end{definition}

%%%%%%%%%%%%%%%%%%%%%%%%%%%%%%%%%%%%%%%%%%%%%%%%%%%%%%%%%%%%%%%%%%%%%%%%%%%%%%%%
% MODELING  --  max k pages (now, n)
%%%%%%%%%%%%%%%%%%%%%%%%%%%%%%%%%%%%%%%%%%%%%%%%%%%%%%%%%%%%%%%%%%%%%%%%%%%%%%%%
\section{Modeling the vehicle-sharing system}\label{sec:modeling}

This section introduces the mathematical models of the vehicle-sharing system. The fleet dynamics describe how users' pick-ups and drop-offs affect the distribution of vehicles. The user behavior model captures how users respond to the incentives offered by the operator. Connecting the two, we then define the incentive-response map. 

\subsection{Vehicle-sharing system dynamics}\label{sec:system_dynamics}

We consider a free-floating vehicle-sharing system where vehicles can be parked anywhere within an operational area divided into regions \cite{Cepolina2014-tm}. We label these using $i \in \mathcal{R}$, where $\mathcal{R} \subset \mathbb{N}$ denotes the set of all regions and $N\!=\!|\mathcal{R}|$ is the number of regions. We assume large parking capacity, so that any amount of vehicles can occupy a region at any time.

We assume the vehicle-sharing system to be instant access, namely, users can take available vehicles at any time, without prior reservation \cite{Cepolina2014-tm}. We divide time into intervals of length $\Delta t$, matching the time resolution of the demand forecasts, and we describe users and vehicles in terms of flows that are constant within each interval. We denote by $x\in \mathbb{R}_+^N$ the vehicle distribution, whose entry $x^i \in \mathbb{R}_+$ represents the number of vehicles available in region $i$ at the beginning of an interval. We assume accurate forecasts of the demand flows $d^{\rmpu,i}, d^{\rmdo,i} \in \mathbb{R}_+$, i.e., the users per hour picking up and dropping off a vehicle in region $i$, respectively, collected in $d^\rmpu \coloneq (d^{\rmpu,i})_{i \in \mathcal{R}}$ and $d^\rmdo \coloneq (d^{\rmdo,i})_{i \in \mathcal{R}}$.

The operator can influence users' choice of pick-up location by offering fare discounts from 0\% to 100\% \cite{Curtale2021-qm}. These incentives $\theta^i \in [0, 1]$ are broadcast to all users in real time, and users picking up a vehicle in region $i$ can redeem the discount. We collect them in $\theta \coloneq (\theta^i)_{i \in \mathcal{R}} \in [0,1]^N$.

Due to these discounts, users located in region $j$ may pick up a vehicle in another region $i \in \mathcal{R}$ instead. We denote by $u^{\rmpu,i} \in \mathbb{R}_+$ the pick-up flow at region $i$, i.e., the number of users per hour picking up vehicles there, and collect these flows in $u^\rmpu \coloneq (u^{\rmpu,i})_{i \in \mathcal{R}} \in \mathbb{R}^N_+$.

Users' drop-off $d^\rmdo$ and pick-up $u^\rmpu$ flows affect the vehicle distribution across regions, according to the dynamics
\begin{align}\label{eq:region_flow_equation}
    x_{t+1} &= x_t+\Delta t \left(d^\rmdo_t-u^\rmpu_t\right), \quad (t \in \mathbb{N})
\end{align}
where the subscript $t$ indicates the time interval. Note that \eqref{eq:region_flow_equation} is the exact discretization of the continuous-time conservation law $\dot{x} = d^\rmdo - u^\rmpu$, since the flows are constant within intervals. Since users can only pick up available vehicles, the pick-up flow must satisfy
\begin{equation}\label{eq:boundary_defs}
    u^\rmpu_t \leq x_t / \Delta t, \quad (\forall t \in \mathbb{N})
\end{equation}
which also ensures that $x_{t+1} \geq 0$.

\subsection{Modeling user behavior}\label{sec:user_behavior}

This section describes how the fare discounts $\theta$ and the vehicle availability $x$ determine the user behavior within each time interval\footnote{Hence, we omit the time index $t$.}. We build our model in two steps: (a) first, we consider the users located in a single region in isolation, and model their choices with a logit model constrained by the availability of vehicles \cite{Atarashi2025-wp}; (b) then, we extend the model to account for users across all regions, who compete to access the same vehicles, by means of a generalized game.

\paragraph{Single region} We first consider the users located in a single region, say $j \in \mathcal{R}$, who request vehicles at rate $d^{\rmpu,j}$. Each of them either picks up a vehicle in some region $i \in \mathcal{R}$, possibly different from $j$, or leaves the service, e.g., in favor of another transportation mode. We collect these options in the choice set $\mathcal{C}=\mathcal{R} \cup \{\emptyset\}$, where $\emptyset$ denotes leaving. We describe the users' choices at the aggregate level by the vector of shares $p^j \in \mathbb{R}^{|\mathcal{C}|}$, indexed by $\mathcal{C}$ with the regions first and $\emptyset$ last. Its entry $p^{j,c}$ is the fraction of these users choosing $c$ in the interval. We model these shares with the constrained logit model \cite{Atarashi2025-wp}%
\begin{subequations}%
\label{eq:logit_problem_single}
\begin{align}
\label{eq:logit_problem_single_cost}
    p^j(x,\theta) = \operatorname*{arg\,max}_{y \in \mathbb{S}} \;
    & y^\top r^j(\theta) - y^\top \log(y)\\
    \text{s.t.} \quad
    & d^{\rmpu,j}\, \Xi \, y \leq x / \Delta t.
    \label{eq:logit_problem_single_cnstr}
\end{align}
\end{subequations}
where $\mathbb{S} \coloneq \left\{p \in \mathbb{R}^{|\mathcal{C}|}\mid p\geq 0, \; \mathbbm{1}^\top p = 1\right\}$ is the probability simplex, $\log(\cdot)$ is applied entry-wise, $\Xi \coloneq \begin{bmatrix} I_{ N} & 0_{N \times 1}\end{bmatrix}$ is a selection matrix removing the leaving component, and $r^j(\theta) \in \mathbb{R}^{|\mathcal{C}|}$ models the incentive-dependent rewards associated with each choice\footnote{A temperature $\tau > 0$ can be included by replacing $r^j$ with $r^j/\tau$; we set $\tau = 1$ in the derivations without loss of generality.}, namely, for all $c \in \mathcal{C}$,
\begin{equation}\label{eq:user_reward_function}
    r^{j,c}(\theta)\coloneq\begin{cases}
        \lambda^{j,c}\, \theta^c - w^{j,c},&\text{if}\quad c\neq \emptyset,\\
        -w^{j,\emptyset},&\text{if}\quad c=\emptyset.
    \end{cases}
\end{equation}
with $\lambda^j \in \mathbb{R}^N_+$ modeling the sensitivity of the users to the incentives in each region, $w^{j,i} \in \mathbb{R}_+$ the base cost\footnote{This depends on the distance between the centers of regions $j$ and $i$.} of picking up a vehicle in region $i$ for users located in region $j$, and $w^{j,\emptyset} \in \mathbb{R}_+$ is the base cost of leaving the service, with $w^j \in \mathbb{R}^{|\mathcal{C}|}_+$. Without \eqref{eq:logit_problem_single_cnstr}, the maximizer of \eqref{eq:logit_problem_single_cost} is the multinomial logit $y \propto \exp(r^j(\theta))$, commonly used to model vehicle-sharing users \cite{Stokkink2021-ck, Di-Febbraro2019-xd, Liu2022-vm}. The optimization problem \eqref{eq:logit_problem_single_cost} is the variational form of the logit model \cite{Anderson1988-rc}, in which the entropy term accounts for unobserved heterogeneity in users' preferences, and $p^j$ are the choice probabilities of a representative user located in region $j$ \cite{Ben-Akiva1985-kw}. For a large number of users, they are also the fractions of these users choosing each option, so that all flows below are deterministic. Hence, $d^{\rmpu,j} \,\Xi\, p^j$ is the pick-up flow generated by these users, and \eqref{eq:logit_problem_single_cnstr} caps it by the available vehicles as in \eqref{eq:boundary_defs}; its multiplier acts as a shadow price that lowers the attractiveness of regions whose vehicles are exhausted \cite{Bell1995-sue}.

\paragraph{Multi-region} In practice, users access the service from all regions simultaneously, and rank the same pick-up choices differently depending on their specific location\footnote{Heterogeneous preferences \eqref{eq:logit_problem_single_cost} arise since incentives $\theta^i$ are regional, and the base costs $w^j$ depend on the distances from region center $j$.}. Thus, we divide users in \textit{user types}, where user type $j\in \mathcal{R}$ denotes users located at region $j$ with pick-up preferences in \eqref{eq:user_reward_function}. In this multi-region scenario, users of type $j$ face the constrained logit model \eqref{eq:logit_problem_single} where the capacities \eqref{eq:logit_problem_single_cnstr} are reduced by the vehicles taken by the other users. This yields the following coupled problems in $p \coloneq (p^j)_{j \in \mathcal{R}}$
\begin{subequations}\label{eq:logit_problem}
\begin{empheq}[left=\forall j \in \mathcal{R}: \;\, \empheqlbrace]{align}
    \max_{p^j \in \mathbb{S}} \quad & \overbrace{(p^j)^\top  r^j(\theta) - (p^j)^\top  \log(p^j)}^{\coloneq J^j(p^j, \theta)} \label{eq:logit_problem_obj}\\
    \mathrm{s.t.} \quad & \sum_{ k \in \mathcal{R}} d^{\rmpu,k} \, \Xi \, p^k \leq x / \Delta t \label{eq:logit_problem_constr_iq}
\end{empheq}
\end{subequations}
This collection of logit models coupled through the resource constraints \eqref{eq:logit_problem_constr_iq} constitutes a generalized game \cite{Facchinei2009-sv}, whose solutions are known as generalized Nash equilibria (GNEs).

The game in \eqref{eq:logit_problem} admits infinitely many GNEs, which differ in how the shortage of vehicles in each region is split among the user types, technically, in the dual variables of the coupling constraints \eqref{eq:logit_problem_constr_iq}. Here, we consider the so-called variational GNEs (v-GNEs), in which all user types share the same dual variables for \eqref{eq:logit_problem_constr_iq}. This models user types that decide simultaneously and face the same shadow prices for the available vehicles \cite{Facchinei2009-sv}.

\begin{remark}[Choice of equilibrium]\label{rem:choice_vgne}
    An equally plausible alternative to the v-GNE is a first-come, first-served model. There, earlier users have priority, but users arrive in random order within an interval. Each user chooses via the logit model among the vehicles still available, and the response is the expected outcome over all arrival orders. Without availability constraints, this expectation coincides with the multinomial logit~\cite{Ben-Akiva1985-kw}. With constraints, characterizing it is still an open question left for future work.
\end{remark}

The v-GNEs are the solutions to the following variational inequality, or generalized equation, parametric in $\theta$ and $x$:
\begin{equation}\label{eq:VI}
    0 \in F(p, \theta) + \mathcal{N}_{\mathcal{Z}(x)}(p),
\end{equation}
where $\mathcal{N}_{\mathcal{Z}(x)}$ is the normal cone of the feasible set of \eqref{eq:logit_problem},
\begin{equation}\label{eq:feasible_set}
    \mathcal{Z}(x) \coloneq \big\{p \in \mathbb{S}^{N} \;\big|\; \textstyle\sum_{k \in \mathcal{R}} d^{\rmpu,k} \, \Xi \, p^k \leq x / \Delta t\big\},
\end{equation}
with $\mathbb{S}^{N} \coloneq \mathbb{S} \times \dots \times \mathbb{S}$, and $F(\cdot, \theta)=-\left(\nabla_{p^j} J^j(p^j,\theta)\right)_{j \in \mathcal{R}}$ is the pseudo-gradient of the game in \eqref{eq:logit_problem}, which reads as
\begin{equation}\label{eq:pseudo_gradient}
    F(p, \theta) = \log(p) + \mathbbm{1} - r(\theta),
\end{equation}
with $\log(\cdot)$ entry-wise and $r(\theta) \coloneq (r^j(\theta))_{j \in \mathcal{R}}$.

Since the objectives in \eqref{eq:logit_problem} are decoupled, the v-GNEs are the maximizers of the social welfare $W(p,\theta) \coloneq \sum_{j \in \mathcal{R}} J^j(p^j,\theta)$ over $\mathcal{Z}(x)$ \cite{Facchinei2011-th}. These are unique for any $x \in \mathbb{R}^N_+$ and $\theta \in [0,1]^N$ (see Lemma~\ref{lem:exist_unique} in Appendix~\ref{app:exist_unique}), yielding the single-valued parameters-to-v-GNE mapping
\begin{equation}\label{eq:psm}
    p^*(x, \theta) = \left\{ p \in \mathbb{S}^N ~ | ~ p \text{ solves } \eqref{eq:VI} \right\}.
\end{equation}

\subsection{Incentive-response map (IRM)}\label{sec:incentive_response_map}

The connection between incentives $\theta$ and pick-up flows $u^\rmpu$, given vehicle availability $x$, is determined by the incentive-response map (IRM):
\begin{equation}\label{eq:incentive_response_map}
    \mathcal{S}(x, \theta) = D \, p^*(x, \theta),
\end{equation}
where the matrix $D \coloneq (d^\rmpu)^\top \otimes I_{|\mathcal{C}|}$ weighs the distributions of each user type $j$ by the corresponding demand flow $d^{\rmpu,j}$. The first $N$ entries of $\mathcal{S}(x, \theta)$ are the pick-up flows in \eqref{eq:region_flow_equation}, i.e., $u^\rmpu = \Xi \, \mathcal{S}(x, \theta)$, and the last is the flow of users leaving the service, $u^{\mathrm{lu}} = \Xi_\emptyset \, \mathcal{S}(x, \theta)$, with $\Xi_\emptyset \coloneq \begin{bmatrix} 0_{1 \times N} & 1 \end{bmatrix}$.

%%%%%%%%%%%%%%%%%%%%%%%%%%%%%%%%%%%%%%%%%%%%%%%%%%%%%%%%%%%%%%%%%%%%%%%%%%%%%%%%
% OPTIMAL CONTROL  --  max k pages (now, n)
%%%%%%%%%%%%%%%%%%%%%%%%%%%%%%%%%%%%%%%%%%%%%%%%%%%%%%%%%%%%%%%%%%%%%%%%%%%%%%%%
\section{Incentive Design Algorithm}\label{sec:optimal_control}

In this section, we formalize the optimal incentive design problem for fleet rebalancing as a mathematical program with nested equilibrium constraints. Then, we develop an efficient first-order algorithm that uses backpropagation and differentiates through the implicit incentive-response maps.

\subsection{Optimal control via incentives}
Fleet rebalancing can be cast as an optimal control problem in which the incentives are the decision variables and the input of the dynamics \eqref{eq:region_flow_equation} is given by the IRM \eqref{eq:incentive_response_map}. Formally, we aim to find a sequence of incentives $\boldsymbol{\theta} \coloneq (\theta_0, \ldots, \theta_{T-1})$ over a planning horizon $\mathbb{T} \coloneq \{0, \ldots, T-1 \}$ that solves
\begin{mini!}|s|
    {\boldsymbol{\theta}}{\sum\limits_{t=0}^{T-1} \ell\left(u_t, \theta_t\right) + h(x_T)\label{eq:inc_plan_prob_obj}}
    {\label{eq:incentive_planning_problem}}
    {}
    \addConstraint{x_{t+1}}{= x_t + \Delta t \left(d^\rmdo_t - \Xi\, u_t\right),\quad \forall t \in \mathbb{T} \label{eq:inc_plan_prob_constr_flow_eq}}
    \addConstraint{u_t}{= \mathcal{S}\left(x_t, \theta_t\right),\quad \forall t \in \mathbb{T} \label{eq:inc_plan_prob_constr_inc_resp}}
    \addConstraint{\theta_t}{\in \Theta, \quad \forall t \in \mathbb{T}, \label{eq:inc_plan_prob_constr_inc_feas}}
\end{mini!}
where $x_0$ is the initial vehicle distribution, $\Theta \coloneq [0,\bar{\theta}]^{N}$ with maximum incentive rate $\bar{\theta} \in (0,1]$, and the stage cost
\begin{equation}\label{eq:stage_cost}
    \ell(u_t, \theta_t) \coloneq \gamma_R \|\theta_t\|_2^2 + \gamma_Q \left(\Delta t \, \Xi_\emptyset\, u_t\right)^2
\end{equation}
trades off the cost of giving incentives against the number of users leaving the service, with $\gamma_R, \gamma_Q \in \mathbb{R}_+$, while the terminal cost  $h(x_T) \coloneq \gamma_h \|x_T - \bar{x}\|_2^2$ penalizes the deviation from a desired final distribution $\bar{x}$ of the fleet, with $\gamma_h \in \mathbb{R}_+$. %Note that the capacity constraints \eqref{eq:boundary_defs}, and hence $x_{t+1} \geq 0$, are enforced by the IRM.

The constraints \eqref{eq:inc_plan_prob_constr_inc_resp} involve the IRM, which is nonlinear, nonsmooth and has no explicit expression, and apply it recursively over the horizon through the dynamics \eqref{eq:inc_plan_prob_constr_flow_eq}, creating a chain of nested equilibrium constraints.

To approach \eqref{eq:incentive_planning_problem}, we first recast it in a more compact form. Substituting \eqref{eq:inc_plan_prob_constr_inc_resp} into the dynamics \eqref{eq:inc_plan_prob_constr_flow_eq} and into the stage cost \eqref{eq:stage_cost} yields the state update map and the stage cost as functions of the state and the incentives only\footnote{Through the known demand $d^\rmpu_t$, $d^\rmdo_t$, the matrix $D$ in~\eqref{eq:incentive_response_map}, and hence $\mathcal{S}$, $\mathcal{T}$, and $g$, also depend on $t$; we omit this dependence for brevity.},
\begin{align}
    \mathcal{T}(x_t, \theta_t) &\coloneq x_t + \Delta t \left(d^\rmdo_t - \Xi\, \mathcal{S}(x_t, \theta_t)\right), \label{eq:state_update_map}\\
    g(x_t, \theta_t) &\coloneq \ell\left(\mathcal{S}(x_t, \theta_t), \theta_t\right). \label{eq:stage_cost_state}
\end{align}
Substituting, in turn, the dynamics into the cost yields 
\begin{subequations}\label{eq:incentive_planning_problem_simplified}
\begin{align}
    \underset{\boldsymbol{\theta}}{\min}\quad
    & V_0\left(x_0,\boldsymbol{\theta}\right) \\
    \text{s.t.}\quad
    & \boldsymbol{\theta}\in\Omega
      \coloneq\Theta\times\dots\times\Theta.
\end{align}
\end{subequations}
in which the constraints \eqref{eq:inc_plan_prob_constr_flow_eq}--\eqref{eq:inc_plan_prob_constr_inc_resp} are embedded in the cost-to-go $V_0$, defined recursively, for all $t \in \mathbb{T}$, as
\begin{equation}\label{eq:cost_to_go}
    V_t(x_t, \boldsymbol{\theta}) \coloneq g(x_t, \theta_t) + V_{t+1}\left(\mathcal{T}(x_t, \theta_t), \boldsymbol{\theta}\right),
\end{equation}
with $V_T(x_T, \boldsymbol{\theta}) \coloneq h(x_T)$; namely, $V_t(x_t, \boldsymbol{\theta})$ is the cost accumulated from interval $t$ to the end of the horizon, starting from $x_t$ under $\boldsymbol{\theta}$. Unrolling~\eqref{eq:cost_to_go} from $t=0$ yields
\begin{equation}
  V_0(x_0,\boldsymbol{\theta}) = \sum_{t=0}^{T-1} g(x_t,\theta_t) + h(x_T),
\end{equation}
where, for all $t \geq 1$,
\begin{equation}
  x_t = \mathcal{T}\Big(\cdots \mathcal{T}\big(\mathcal{T}(x_0,\theta_0),\theta_1\big) \cdots,\theta_{t-1}\Big),
\end{equation}
which reveals the nested structure of~\eqref{eq:incentive_planning_problem_simplified}.

\subsection{Algorithm design and analysis}
To solve~\eqref{eq:incentive_planning_problem_simplified}, we use the first-order scheme for nonsmooth functions summarized in Alg.~\ref{alg:proj_grad_desc}. Computing a ``gradient'' of $V_0$ is complicated (i) by the IRM~\eqref{eq:incentive_response_map}, which is implicit and nonsmooth, and (ii) by its nested application along the horizon in~\eqref{eq:incentive_planning_problem}. To solve these challenges, we adapt the algorithm in~\cite{Zuliani2023-po} to our game-theoretic setup with nonlinear cost functions. Namely, we address (i) by relying on conservative Jacobians (Def.~\ref{def:conservative_jacobian}), computed via nonsmooth implicit differentiation~\cite{Bolte2021-ml, Bolte2022-uo}, and (ii) by propagating these Jacobians backward in time along the recursion~\eqref{eq:cost_to_go} (Alg.~\ref{alg:bwrd_prop}--\ref{alg:fwrd_pass}). The remainder of this section presents these algorithms and establishes convergence guarantees (Thm.~\ref{thm:conv_to_crit_point}).

\begin{algorithm}
    \caption{Projected conservative-gradient update}\label{alg:proj_grad_desc}
    \begin{algorithmic}[1]
        \Require $x_0$, $\boldsymbol{\theta}^{(0)}$, $\{\alpha^{(k)}\}_{k \in \mathbb{N}}$, $\varepsilon$
        \State $k \gets 0$
        \Repeat
            \State $J_{V,\boldsymbol{\theta}}^{(k)} \in \mathcal{J}_{\boldsymbol{\theta}} V_0(x_0, \boldsymbol{\theta}^{(k)})$ \Comment{computed via $\AlgRef{alg:bwrd_prop}{x_0, \boldsymbol{\theta}^{(k)}}$}\label{line:pgd_jacobian}
            \State $\boldsymbol{\theta}^{(k+1)} \gets \Pi_{\Omega}\big(\boldsymbol{\theta}^{(k)} - \alpha^{(k)} J_{V,\boldsymbol{\theta}}^{(k)}\big)$ \Comment{Projected gradient step}\label{line:pgd_step}
            \State $k \gets k+1$
        \Until{$\|\boldsymbol{\theta}^{(k)} - \boldsymbol{\theta}^{(k-1)}\|_2 < \varepsilon$}
        \State \Return $\boldsymbol{\theta}^{(k)}$
    \end{algorithmic}
\end{algorithm}

To compute a conservative gradient $J_{V,\boldsymbol{\theta}}^{(k)} \in \mathcal{J}_{\boldsymbol{\theta}} V_0(x_0, \boldsymbol{\theta}^{(k)})$ in line~\ref{line:pgd_jacobian} of Alg.~\ref{alg:proj_grad_desc}, we leverage the recursive structure of the cost-to-go in \eqref{eq:cost_to_go}. Applying the chain rule for conservative Jacobians~\cite[Lemma~5]{Bolte2021-cs} to~\eqref{eq:cost_to_go} expresses $J^{(t)}_{V,x} \in \mathcal{J}_x V_t(x_t, \boldsymbol{\theta})$ and $J^{(t)}_{V,\boldsymbol{\theta}} \in \mathcal{J}_{\boldsymbol{\theta}} V_t(x_t, \boldsymbol{\theta})$ recursively in terms of their counterparts at $t+1$ (Alg.~\ref{alg:bwrd_prop}, lines~\ref{line:bp_Vx}--\ref{line:bp_Vtheta}). This recursion also involves the partial Jacobians of the stage cost $g$ in~\eqref{eq:stage_cost_state} and of the state update map $\mathcal{T}$ in~\eqref{eq:state_update_map} at $(x_t, \theta_t)$. It evaluates this recursion backwards in time from $t = T$, where $V_T(x_T, \boldsymbol{\theta}) = h(x_T)$ by definition  (line~\ref{line:bp_init_x}), down to $t = 0$, where $J^{(0)}_{V,\boldsymbol{\theta}}$ is the sought conservative gradient of $V_0$.

\begin{algorithm}[t]
    \caption{Backpropagation}\label{alg:bwrd_prop}
    \begin{algorithmic}[1]
        \Require $x_0$, $\boldsymbol{\theta} = (\theta_0, \dots, \theta_{T-1})$
        \State $\boldsymbol{x} \gets$ state trajectory of~\eqref{eq:state_update_map} under $\boldsymbol{\theta}$ \Comment{via $\AlgRef{alg:fwrd_pass}{x_0, \boldsymbol{\theta}}$}\label{line:bp_forward}
        \State $J_{V,x}^{(T)} \gets 2\gamma_h (x_T - \bar{x})^\top,\; J_{V,\boldsymbol{\theta}}^{(T)} \gets 0$ \Comment{cons. Jac. of $V_T = h$}\label{line:bp_init_x}
        \For{$t = T-1, \dots, 0$}
            \State $(J_{g,x}^{(t)}, J_{g,\boldsymbol{\theta}}^{(t)}) \in \mathcal{J} g(x_t, \theta_t)$ \Comment{computed via~\eqref{eq:jac_g}}\label{line:bp_impl_diff}
            \State $(J_{\mathcal{T},x}^{(t)}, J_{\mathcal{T},\boldsymbol{\theta}}^{(t)} ) \in \mathcal{J} \mathcal{T}(x_t, \theta_t)$  \Comment{computed via~\eqref{eq:jac_T}}\label{line:bp_JT}
            \State $J_{V,x}^{(t)} \gets J_{g,x}^{(t)} + J_{V,x}^{(t+1)} J_{\mathcal{T},x}^{(t)}$ \Comment{$\in \mathcal{J}_x V_t$}\label{line:bp_Vx}
            \State $J_{V,\boldsymbol{\theta}}^{(t)} \gets J_{g,\boldsymbol{\theta}}^{(t)} + J_{V,x}^{(t+1)} J_{\mathcal{T},\boldsymbol{\theta}}^{(t)} + J_{V,\boldsymbol{\theta}}^{(t+1)}$ \Comment{$\in \mathcal{J}_{\boldsymbol{\theta}} V_t$}\label{line:bp_Vtheta}
        \EndFor
        \State \Return $J_{V,\boldsymbol{\theta}}^{(0)}$ \Comment{a conservative gradient of $V_0(x_0, \boldsymbol{\theta})$}
    \end{algorithmic}
\end{algorithm}

\begin{algorithm}
    \caption{Forward pass}\label{alg:fwrd_pass}
    \begin{algorithmic}[1]
        \Require $x_0$, $\boldsymbol{\theta} = (\theta_0, \dots, \theta_{T-1})$
        \For{$t = 0, \dots, T-1$}
            \State $p^*_t \gets \SolRef{eq:VI}[x_t, \theta_t]$ \Comment{v-GNE of \eqref{eq:logit_problem}}\label{line:fp_vgne}
            \State $x_{t+1} \gets x_t + \Delta t \left(d^\rmdo_t - \Xi D\, p^*_t\right)$ \Comment{state update~\eqref{eq:state_update_map}}\label{line:fp_step}
        \EndFor
        \State \Return $\boldsymbol{x}=(x_0, \dots, x_T)$ \Comment{state trajectory from $x_0$ under $\boldsymbol{\theta}$}
    \end{algorithmic}
\end{algorithm}

Implementing this recursion requires (i) the state trajectory $\boldsymbol{x}^{(k)}=(x_0, x_1^{(k)}, \dots, x_T^{(k)})$ generated by the control sequence $\boldsymbol{\theta}^{(k)}$ from the initial state $x_0$, computed by the so-called forward pass in Alg.~\ref{alg:fwrd_pass} (line~\ref{line:bp_forward}), as well as (ii) the partial Jacobians of the state update map~\eqref{eq:state_update_map} and of the stage cost~\eqref{eq:stage_cost_state} at $(x_t, \theta_t)$. Both depend on $(x_t, \theta_t)$ through the IRM, i.e., through the v-GNE mapping $p^*$ in~\eqref{eq:psm}. Given $J^{(t)}_{p^*,x} \in \mathcal{J}_x p^*(x_t, \theta_t)$ and $J^{(t)}_{p^*,\theta} \in \mathcal{J}_\theta p^*(x_t, \theta_t)$, the chain rule applied to~\eqref{eq:incentive_response_map} and~\eqref{eq:stage_cost}--\eqref{eq:stage_cost_state} yields
\begin{subequations}\label{eq:jac_g}
\begin{align}
    J^{(t)}_{g,x} &= c_t\, J^{(t)}_{p^*,x}, \label{eq:jac_g_x}\\
    J^{(t)}_{g,\boldsymbol{\theta}} &= e_t^\top \otimes \big(2\gamma_R\, \theta_t^\top + c_t\, J^{(t)}_{p^*,\theta}\big), \label{eq:jac_g_theta}
\end{align}
\end{subequations}
and
\begin{subequations}\label{eq:jac_T}
\begin{align}
    J^{(t)}_{\mathcal{T},x} &= I - \Delta t\, \Xi D\, J^{(t)}_{p^*,x}, \label{eq:jac_T_x}\\
    J^{(t)}_{\mathcal{T},\boldsymbol{\theta}} &= e_t^\top \otimes \big(-\Delta t\, \Xi D\, J^{(t)}_{p^*,\theta}\big), \label{eq:jac_T_theta}
\end{align}
\end{subequations}
where $c_t \coloneq 2\gamma_Q \Delta t^2\, u^{\mathrm{lu}}_t\, \Xi_\emptyset D$, with $u^{\mathrm{lu}}_t = \Xi_\emptyset D\, p^*(x_t, \theta_t)$, is the Jacobian of the leaving-user term in~\eqref{eq:stage_cost} with respect to $p$. Since $\mathcal{T}$ and $g$ at interval $t$ depend on $\boldsymbol{\theta}$ only through $\theta_t$, their Jacobians with respect to $\boldsymbol{\theta}$ vanish outside the block of $\theta_t$, which we encode via $e_t \in \mathbb{R}^T$.

It remains to compute $J^{(t)}_{p^*,x}$ and $J^{(t)}_{p^*,\theta}$. Since $p^*$ is defined only implicitly, we use the nonsmooth implicit function theorem (IFT)~\cite{Bolte2021-ml}. Applied directly to~\eqref{eq:VI}, e.g., via the fixed-point form $p = \Pi_{\mathcal{Z}(x)}\big(p - F(p,\theta)\big)$, the IFT would require a conservative Jacobian of the projection onto $\mathcal{Z}(x)$, which depends on $x$ and has no closed form. Instead, we adapt the dual route of~\cite{Zuliani2023-po} from quadratic to entropic costs. This route avoids differentiating the projection onto $\mathcal{Z}(x)$ by working with the dual problem in the multipliers $\mu = (\mu_\rmeq, \mu_\rmiq)$ of the equality and inequality constraints of $\mathcal{Z}(x)$. Its feasible set $\mathcal{M} \coloneq \mathbb{R}^N \times \mathbb{R}^N_{\geq 0}$ is a box, independent of $x$ and $\theta$, so the corresponding projection reduces to the entry-wise $\Pi_\mathcal{M}(\mu) = \big(\mu_\rmeq, \max(0, \mu_\rmiq)\big)$, whose conservative Jacobian is explicit. Appendix~\ref{app:vgne_sensitivity} details the derivation and summarizes the procedure in Alg.~\ref{alg:impl_diff}.

Next, we show that the accumulation points of Algorithm~\ref{alg:proj_grad_desc} are critical points of~\eqref{eq:incentive_planning_problem_simplified}. This guarantee does not follow directly from~\cite{Zuliani2023-po}, whose analysis covers nested optimization problems with quadratic costs. In contrast, \eqref{eq:incentive_planning_problem} involves nested v-GNEs, whose equivalent social-welfare problems have entropic, hence non-quadratic, costs. Nevertheless, the analysis of~\cite{Zuliani2023-po} can be adapted to our setting by exploiting the path-differentiability of the v-GNE mapping $p^*$.

\begin{theorem}\label{thm:conv_to_crit_point}
    Let $\boldsymbol{\theta}^{(0)} \in \Omega$, and let the step sizes satisfy $\alpha^{(k)} > 0$ for all $k \in \mathbb{N}$, $\sum_{k=0}^\infty \alpha^{(k)} = \infty$, and $\sum_{k=0}^\infty (\alpha^{(k)})^2 < \infty$. Assume that, for some $\delta > 0$, the state trajectories $\boldsymbol{x}^{(k)}$ computed by Algorithm~\ref{alg:fwrd_pass} along the iterations of Algorithm~\ref{alg:proj_grad_desc} satisfy $x^{(k)}_t \geq \delta \mathbbm{1}$ for all $t \in \{0, \dots, T\}$ and $k \in \mathbb{N}$. Then, every accumulation point $\tilde{\boldsymbol{\theta}}$ of $\{\boldsymbol{\theta}^{(k)}\}_{k \in \mathbb{N}}$ is a critical point of~\eqref{eq:incentive_planning_problem_simplified}, i.e., $0 \in \mathcal{J}_{\boldsymbol{\theta}} V_0(x_0, \tilde{\boldsymbol{\theta}}) + \mathcal{N}_\Omega(\tilde{\boldsymbol{\theta}})$, and the sequence $\{V_0(x_0, \boldsymbol{\theta}^{(k)})\}_{k \in \mathbb{N}}$ converges.
    {\hfill $\square$}
\end{theorem}

\noindent The proof follows arguments analogous to~\cite{Zuliani2023-po}, adapted to lower-level problems with entropic cost functions as in~\eqref{eq:VI}, and \ifarxiv is given in Appendix~\ref{sec:apx_guarantee_proof}.\else is omitted due to space limitations. For completeness, we provide it in the extended version of this paper~\cite[App.~C]{tihanyi2026extended}.\fi 

\begin{remark}[Non-empty regions]\label{rem:nonempty_regions}
    The assumption $x^{(k)}_t \geq \delta \mathbbm{1}$ in Theorem~\ref{thm:conv_to_crit_point} ensures that the dual route of Alg.~\ref{alg:impl_diff} is well defined along the iterates. It plays the role of the assumption in~\cite[Thm.~2]{Zuliani2023-po}. There, the lower-level problems are strongly convex quadratic programs, and requiring them to remain feasible and satisfy LICQ along the iterates suffices for the dual route. In our setting, $\mathcal{Z}(x)$ is always nonempty and satisfies LICQ (Appendix~\ref{app:vgne_sensitivity}), yet this is not enough: if a region runs empty, i.e., $x^i_t = 0$, the dual minimum is not attained. We therefore require strict feasibility, namely that no region runs empty, with a uniform margin $\delta$ along the iterates. The margin also keeps the iterates in a compact set on which $V_0(x_0,\cdot)$ is path-differentiable. The assumption holds, e.g., if $x_0 > 0$ and $d^\rmdo_t > 0$ for all $t \in \mathbb{T}$, since then $x_{t+1} \geq \Delta t\, d^\rmdo_t$ by~\eqref{eq:state_update_map}. Otherwise, it can be verified a posteriori on the trajectories computed by Alg.~\ref{alg:fwrd_pass}.
\end{remark}

%%%%%%%%%%%%%%%%%%%%%%%%%%%%%%%%%%%%%%%%%%%%%%%%%%%%%%%%%%%%%%%%%%%%%%%%%%%%%%%%
% RESULTS  --  max k pages (now n)
%%%%%%%%%%%%%%%%%%%%%%%%%%%%%%%%%%%%%%%%%%%%%%%%%%%%%%%%%%%%%%%%%%%%%%%%%%%%%%%%
\section{Numerical results}\label{sec:numerical_results}

We evaluate our approach on the Helsinki Region Transport (HSL) city-bike trip data~\cite{hsl2025od}. Since the system is station-based, we aggregate the stations into $N=46$ regions by agglomerative clustering with a 300~m threshold (Fig.~\ref{fig:map_region_bounds}), and treat each region as free-floating with large parking capacity. We discretize time into $T=96$ 15-minute steps over May 1, 2025, a public holiday with about twice the trips of a regular day, i.e., a peak-demand scenario. We set $\lambda^{j,i} = 500$ for all region pairs, so a full incentive ($\theta^i=1$) compensates 500~m of walking, beyond which most users are unwilling to walk to an alternative vehicle~\cite{Herrmann2014-it}. As neighboring regions are $604$~m apart on average, high enough incentives can often draw users to a nearby region. We set $w^{j,\emptyset} = 100$, i.e., without incentives, users tolerate walking up to 100~m before leaving the service. As even the closest regions are $278$~m apart, users prefer leaving over switching regions absent incentives, an unforgiving baseline. As rewards are expressed in meters, we divide them by the temperature $\tau = 50$~m, which sets how sharply users discriminate between choices. We compute the v-GNEs with CVXPY~\cite{diamond2016cvxpy} using the SCS solver~\cite{odonoghue2016conic}, and set $\gamma_R = 0.5$, $\gamma_Q = 5$, $\gamma_h = 0$ (no terminal cost), $\bar{\theta} = 1$, and $x_0 = 10 \cdot \mathbbm{1}$.

\begin{figure}
    \centering
    \includegraphics[width=0.70\linewidth]{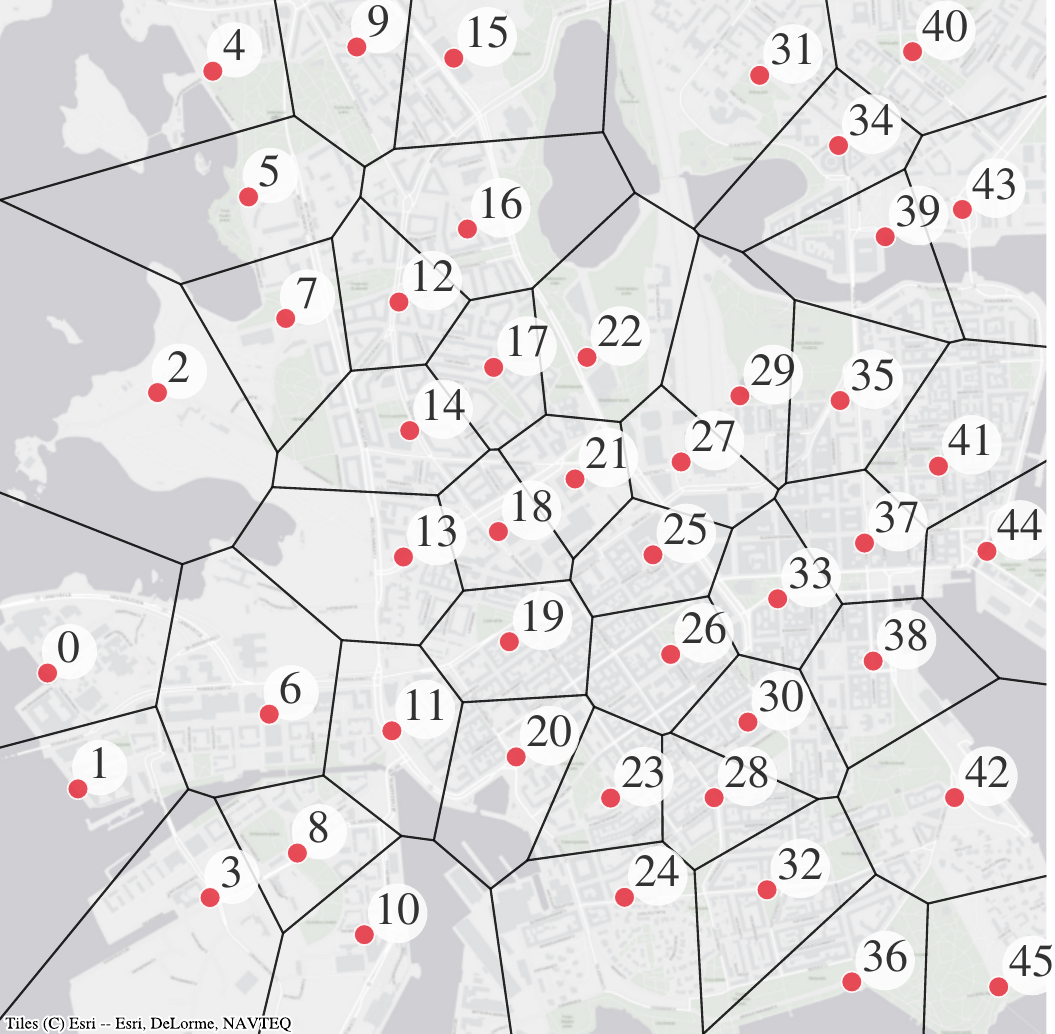}
    \caption{City-bike stations of central Helsinki~\cite{hsl2025od} aggregated into $N=46$ free-floating regions (black boundaries) with indexed centers (red dots). Neighboring centers are $278$ to $2119$~m apart (average $604$~m). Basemap \textcopyright{} 2026 Esri and its licensors; sources: Esri, HERE, Garmin, \textcopyright{} OpenStreetMap contributors, and the GIS user community.}
    \label{fig:map_region_bounds}
\end{figure}

\subsection{Reduction in leaving users}

On this peak-demand day, our optimized incentives reduce the number of leaving users by 61.8$\%$. As pick-up demand exceeds the available vehicles during peak hours (11 a.m.--8 p.m.), the algorithm increases incentives, reducing the number of users abandoning the service from midday onward (Fig.~\ref{fig:case_study_results_simple}). This effect holds across 10 consecutive days (May 1--10, 2025), with reductions from 61.8$\%$ on the high-demand holiday (Thursday, May 1) to 98.8$\%$ on a low-demand Sunday (May 4), where the zero-incentive baseline has about a third of the leaving users (Fig.~\ref{fig:case_study_results_simple_days}).

\begin{figure}
    \centering
    \includegraphics[width=0.99\linewidth]{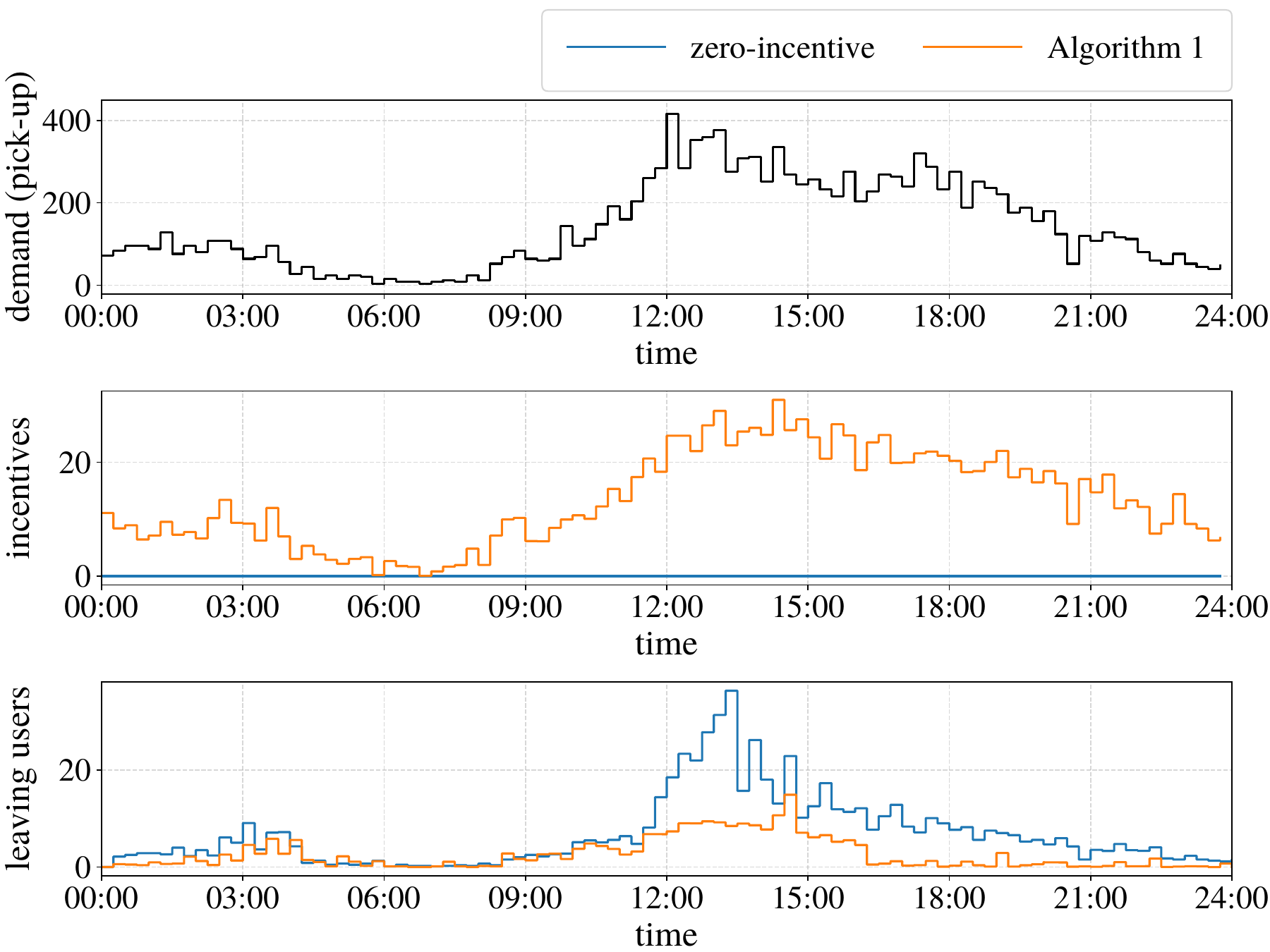}
    \caption{Incentive trajectory optimized by Algorithm~\ref{alg:proj_grad_desc} against the zero-incentive baseline. All plots show aggregated flow across all regions against the time of day. Top: aggregated pick-up demand flow, $\sum_{i \in \mathcal{R}} d_t^{\rmpu,i}$. Center: aggregated incentive trajectories, $\sum_{i \in \mathcal{R}} \theta_t^i$, generated by Algorithm~\ref{alg:proj_grad_desc} (orange). Bottom: flow of users leaving the service, $u^{\mathrm{lu}}_t$, without incentives (blue) and with incentives generated by Algorithm~\ref{alg:proj_grad_desc} (orange).}
    \label{fig:case_study_results_simple}
\end{figure}

\begin{figure}
    \centering
    \includegraphics[width=0.99\linewidth]{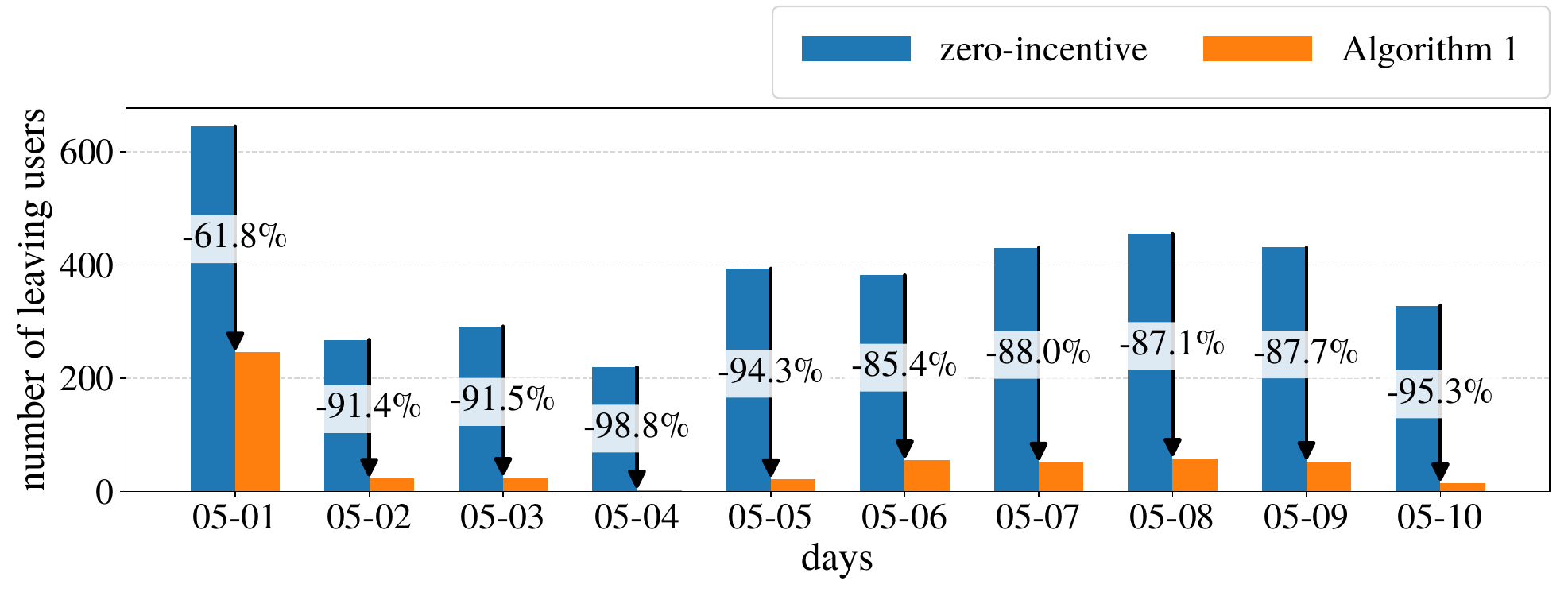}
    \caption{Daily number of leaving users without incentives (blue) and with the incentives computed by Algorithm~\ref{alg:proj_grad_desc} (orange) over ten consecutive days (1--10 May 2025). Labels report the relative reduction.}
    \label{fig:case_study_results_simple_days}
\end{figure}

We compare Algorithm~\ref{alg:proj_grad_desc} against the zero-incentive baseline and three naive alternatives: i) Trigger, a threshold-triggered policy, where $\theta_t^i=\bar{\theta}$, if there are more vehicles in region $i$ than threshold $P_\mathrm{trig.}$, $x_t^i\geq P_\mathrm{trig.}$, and $\theta_t^i=0$ otherwise, with $P_\mathrm{trig.} = 10$; ii) Proportional, a proportional-error policy, where $\theta_t^i=P_\mathrm{prop.} \cdot (x_t^i-d_t^{\rmpu,i})$, the incentives are set proportional to the overshoot of available vehicles over the demanded vehicles, with $P_\mathrm{prop.} = 0.1$; iii) One-step, a short-horizon policy, where $\theta_t$ is generated by solving \eqref{eq:incentive_planning_problem} for a single-step horizon $T=1$. We summarized the results in Table~\ref{tab:algorithm_comp}. Algorithm~\ref{alg:proj_grad_desc} achieves the largest reduction in leaving users (61.8$\%$, against 41.5$\%$--59.3$\%$ for the baselines) while using the least incentive, yielding $1.1\times$ to $2.7\times$ the reduction per incentive.

\begin{table}[t]
    \centering
    \begin{tabular}{|c||c|c|c|c|}
        \hline
        Algorithms & \shortstack{Leaving\\ users} &  Reduction & \shortstack{Incentive\\ usage} & \shortstack{Incentive\\ efficiency} \\
        \hline\hline
        zero-incentive & 645 & - & - & - \\
        Algorithm~\ref{alg:proj_grad_desc} & 246 & -61.8$\%$ & 1280 & -0.048$\%$\\
        trigger & 377 & -41.5$\%$ & 1647 & -0.025$\%$\\
        proportional & 337 & -47.7$\%$ & 2690 & -0.018$\%$\\
        one-step & 262 & -59.3$\%$ & 1389 & -0.043$\%$\\
        \hline
    \end{tabular}
    
    \vspace{\baselineskip}
    \refstepcounter{table}\label{tab:algorithm_comp}%
    \parbox[t]{\hsize}{\footnotesize\noindent TABLE~\thetable.~~Comparison of the five applied algorithms: i) Leaving users -- the total number of leaving users throughout the day across all regions; ii) Reduction -- improvement in the reduction in leaving user numbers compared to the zero-incentive baseline; iii) Incentive usage -- total amount of incentives broadcast throughout the day across all regions; iv) Incentive efficiency -- reduction divided by incentive usage.}
\end{table}

\subsection{Convergence behavior}

Since \eqref{eq:incentive_planning_problem_simplified} is non-convex, different initial guesses may converge to different critical points. To assess the practical impact of this, we ran 100 experiments with initial guesses $\boldsymbol{\theta}^{(0)}$ drawn uniformly from $\Omega$ and tracked the cost function value across iterations for each run (Fig.~\ref{fig:convergence_rand_init}). Across all 100 randomized runs, the algorithm shows the same convergence behavior: the median cost decreases by about $70\%$ and plateaus after roughly 12 iterations. After 20 iterations with $\alpha^{(k)} = (k+1)^{-3}$, the range of the costs across all runs is $29.5\%$ of the median, indicating that the critical points reached from different initializations have comparable performance. This step size decays faster than required by Theorem~\ref{thm:conv_to_crit_point} and favors fast practical progress.

\begin{figure}
    \centering
    \includegraphics[width=0.99\linewidth]{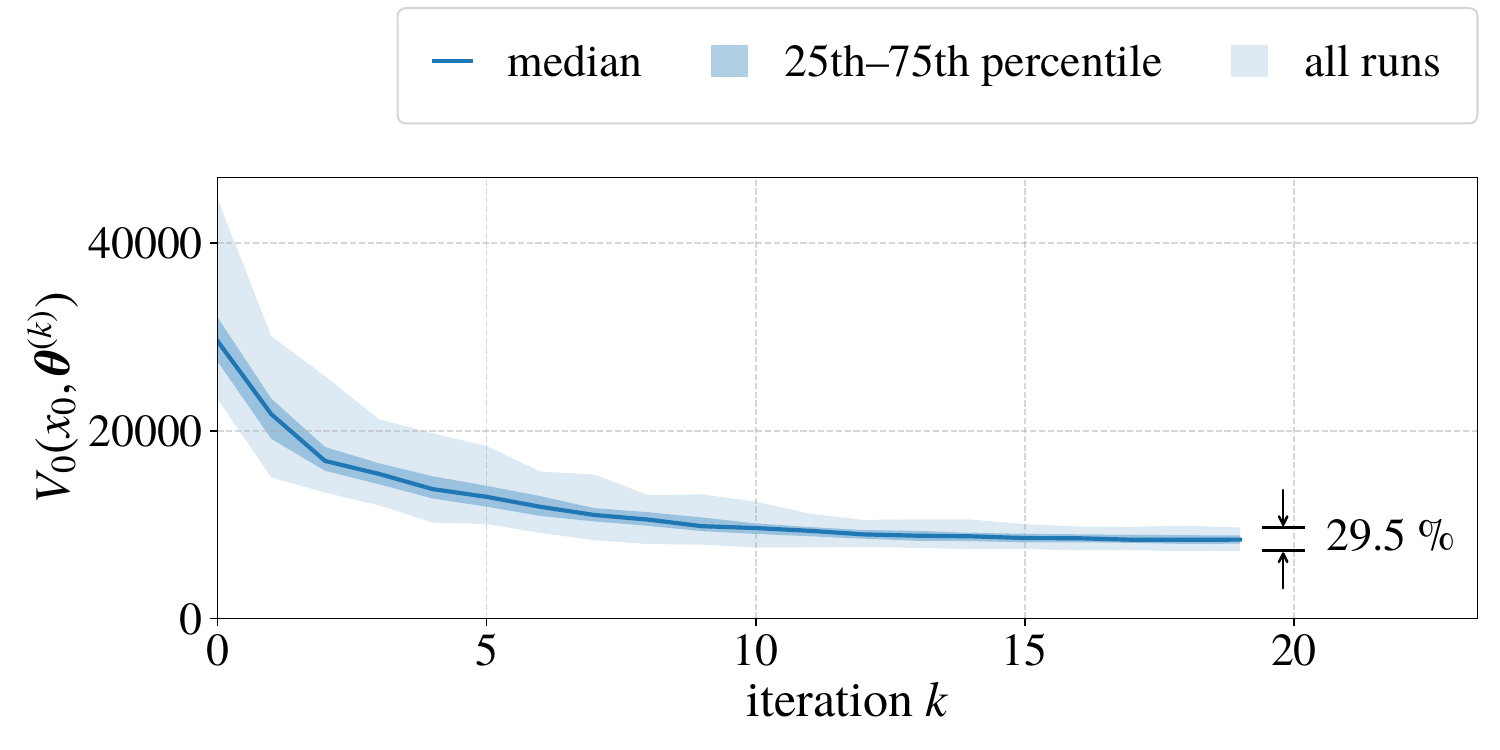}
    \caption{Cost $V_0(x_0, \boldsymbol{\theta}^{(k)})$ per iteration of Algorithm~\ref{alg:proj_grad_desc} for 100 random $\boldsymbol{\theta}^{(0)} \in \Omega$: median (line), 25th--75th percentile (dark band) and range of all runs (light band). After 20 iterations, the range is $29.5\%$ of the median.}
    \label{fig:convergence_rand_init}
\end{figure}

%%%%%%%%%%%%%%%%%%%%%%%%%%%%%%%%%%%%%%%%%%%%%%%%%%%%%%%%%%%%%%%%%%%%%%%%%%%%%%%%
% CONCLUSION  --  max k pages (now, n)
%%%%%%%%%%%%%%%%%%%%%%%%%%%%%%%%%%%%%%%%%%%%%%%%%%%%%%%%%%%%%%%%%%%%%%%%%%%%%%%%
\section{Conclusion}\label{sec:conclusion}

We formulated incentive-based fleet rebalancing as an optimal control problem whose inputs are the equilibrium responses of competing users, modeled as v-GNEs of availability-constrained logit models. Differentiating these equilibria through their dual fixed points yields a first-order method that backpropagates through the day and whose accumulation points are critical points. On Helsinki bike-sharing data, the resulting incentives cut the number of leaving users by 62--99\% and outperformed heuristic and myopic policies with the lowest incentive usage.

Several limitations point to future work. Incentives act on pick-ups only, thus drop-off incentives are an immediate extension, as a logit model of the return location would replace the exogenous drop-off flows without additional coupling. Region-wide discounts also reward users who would have picked up anyway, which type-dependent incentives would avoid. The v-GNE is one of several plausible equilibrium selections (Remark~\ref{rem:choice_vgne}), and the choice model awaits calibration on field data. Finally, the open-loop plan assumes perfect demand forecasts, and a receding-horizon implementation would hedge against forecast errors.

%%%%%%%%%%%%%%%%%%%%%%%%%%%%%%%%%%%%%%%%%%%%%%%%%%%%%%%%%%%%%%%%%%%%%%%%%%%%%%%%
% APPENDIX  --  max k pages (now, n)
%%%%%%%%%%%%%%%%%%%%%%%%%%%%%%%%%%%%%%%%%%%%%%%%%%%%%%%%%%%%%%%%%%%%%%%%%%%%%%%%
\section{Appendix}\label{sec:appendix}

\subsection{Existence and uniqueness of the v-GNE}\label{app:exist_unique}

\begin{lemma}\label{lem:exist_unique}
    For any $x \in \mathbb{R}^N_+$, $\theta \in [0,1]^N$, the generalized equation in \eqref{eq:VI} admits a unique solution, $p^*(x, \theta)$.
    {\hfill $\square$}
\end{lemma}
\begin{proof}
  \emph{Existence.} For any $x \in \mathbb{R}^N_+$, $\mathcal{Z}(x)$ is compact and nonempty, as it contains the profile $\hat{p}$ in which all users leave, i.e., $\hat{p}^{j,\emptyset}=1$ and $\hat{p}^{j,i}=0$ for all $j \in \mathcal{R}$ and $i \neq \emptyset$. Hence, the maximizer exists. \emph{Uniqueness.} The Hessian of the social welfare is $\nabla^2_p W(p,\theta) = -\rmdiag(p)^{-1} \prec 0$ for all $p > 0$. Thus, $W(\cdot,\theta)$ is strictly concave on the interior of $\mathbb{S}^N$, and, by continuity of $t \mapsto t \log t$ on $[0,\infty)$ with $0 \log 0 \coloneqq 0$, on all of $\mathbb{S}^N$. Therefore, the maximizer is unique.
\end{proof}

\subsection{Conservative Jacobians of the v-GNE}\label{app:vgne_sensitivity}

This appendix details the computation of the conservative Jacobians $J_{p^*,x} \in \mathcal{J}_x p^*(x, \theta)$ and $J_{p^*,\theta} \in \mathcal{J}_\theta p^*(x, \theta)$ of the v-GNE mapping~\eqref{eq:psm}, required in~\eqref{eq:jac_g}--\eqref{eq:jac_T}. The procedure is summarized in Alg.~\ref{alg:impl_diff}; we omit the time index $t$ throughout.

\begin{algorithm}
    \caption{Conservative Jacobians of the v-GNE}\label{alg:impl_diff}
    \begin{algorithmic}[1]
        \Require $x$, $\theta$
        \AlgStep{Solving the v-GNE and its dual}{step:sol_lower_level_prob}
        \State $p^*, \mu^* \gets \SolRef{eq:VI}[x, \theta]$ \Comment{Solve~\eqref{eq:VI} and its dual}\label{line:id_solve}
        \State $P^* \gets \rmdiag(p^*)$ \Comment{Shorthand}\label{line:id_P}
        \AlgStep{Differentiating the projection onto $\mathcal{M}$}{step:diff_proj_op}
        \State $\mu' = (\mu'_\rmeq, \mu'_\rmiq) \gets \mu^* + G p^* - b(x)$ \Comment{Argument in~\eqref{eq:fixed_point}}\label{line:id_mu_prime}
        \State $\Sigma \gets \rmdiag(\mathbbm{1}_N, \sigma(\mu'_\rmiq))$ \Comment{Diff.\ $\Pi_\mathcal{M}$, see~\eqref{eq:step_function}}\label{line:id_Sigma}
        \AlgStep{Implicit differentiation of the dual solution}{step:impl_func_thm}
        \State $J_{\mathcal{F},\mu} \gets \Sigma (I - G P^* G^\top) - I$ \Comment{Diff.~\eqref{eq:fixed_point} in $\mu$}\label{line:id_U}
        \State $J_{\mathcal{F},x} \gets -\tfrac{1}{\Delta t} \Sigma E$, \, $J_{\mathcal{F},\theta} \gets \Sigma G P^* \Lambda$ \Comment{Diff.~\eqref{eq:fixed_point}}\label{line:id_V}
        \State $J_{\mu^*,x} \gets -J_{\mathcal{F},\mu}^{-1} J_{\mathcal{F},x}$ \Comment{IFT~\cite[Thm.~5]{Bolte2021-ml}}\label{line:id_Jmu}
        \State $J_{\mu^*,\theta} \gets -J_{\mathcal{F},\mu}^{-1} J_{\mathcal{F},\theta}$ \Comment{IFT~\cite[Thm.~5]{Bolte2021-ml}}\label{line:id_Jmu_theta}
        \AlgStep{Differentiating the primal solution}{step:primal_sol_jacs}
        \State $J_{p^*,x} \gets -P^* G^\top J_{\mu^*,x}$ \Comment{Diff.~\eqref{eq:dual_to_primal}}\label{line:id_Jpx}
        \State $J_{p^*,\theta} \gets P^* (\Lambda - G^\top J_{\mu^*,\theta})$ \Comment{Diff.~\eqref{eq:dual_to_primal}}\label{line:id_Jptheta}
        \State \Return $J_{p^*,x}, J_{p^*,\theta}$
    \end{algorithmic}
\end{algorithm}

\textit{Step~\ref{step:sol_lower_level_prob} (Dual of the social welfare maximization).} We write the constraints of $\mathcal{Z}(x)$ in~\eqref{eq:feasible_set} compactly as $G_\rmeq\, p = \mathbbm{1}_N$ and $G_\rmiq\, p \leq x/\Delta t$, with $G_\rmeq \coloneq I_N \otimes \mathbbm{1}_{|\mathcal{C}|}^\top$ and $G_\rmiq \coloneq \Xi\, D$, and stack them as $G \coloneq (G_\rmeq, G_\rmiq)$ and $b(x) \coloneq (\mathbbm{1}_N, x/\Delta t)$. Let $\mu \in \mathcal{M}$ be the corresponding multipliers. Since the v-GNE maximizes the social welfare $W(\cdot, \theta)$ over $\mathcal{Z}(x)$ (Section~\ref{sec:user_behavior}), stationarity of the Lagrangian yields
\begin{equation}\label{eq:dual_to_primal}
    p^*(x,\theta) = p(\mu^*,\theta) \coloneq \exp\left(r(\theta) - \mathbbm{1} - G^\top \mu^*\right),
\end{equation}
where $r(\theta) = \Lambda \theta - w$, with $\Lambda \coloneq (\Lambda^j)_{j \in \mathcal{R}}$, $\Lambda^j \coloneq \big(\rmdiag(\lambda^j), 0_{1 \times N}\big)$, and $w \coloneq (w^j)_{j \in \mathcal{R}}$, and where $\mu^* = \mu^*(x,\theta)$ minimizes the dual function $\mathbbm{1}^\top p(\mu,\theta) + b(x)^\top \mu$ over $\mathcal{M}$. Unlike in the quadratic case of~\cite{Zuliani2023-po}, where it is affine, this map is an entry-wise exponential. By~\eqref{eq:dual_to_primal}, $p^* > 0$, hence $P^* \coloneq \rmdiag(p^*) \succ 0$. $G$ has full row rank whenever $d^\rmpu \neq 0$: the $\emptyset$-entry of block $j$ appears only in row $j$ of $G_\rmeq$, and the rows of $G_\rmiq$ are nonzero with disjoint supports. Hence, LICQ holds and $\mu^*$ is unique.

\textit{Step~\ref{step:diff_proj_op} (Fixed-point characterization).} The dual minimizer $\mu^*$ is a fixed point of a projected gradient step
\begin{equation}\label{eq:fixed_point}
    0 = \mathcal{F}(\mu^*, x, \theta) \coloneq \Pi_\mathcal{M}\big(\mu^* + G\, p(\mu^*,\theta) - b(x)\big) - \mu^*,
\end{equation}
where the parameters enter $\mathcal{F}$ only through $b(x)$ and $r(\theta)$. The map $\mathcal{F}$ is a composition of affine maps, the entry-wise exponential, and $\max(0,\cdot)$; hence, it is locally Lipschitz and definable in the o-minimal structure $\mathbb{R}_{\exp}$, and thus path-differentiable~\cite{Bolte2021-ml}. Its only nonsmooth component is $\Pi_\mathcal{M}$, whose conservative Jacobian at $\mu'$ (line~\ref{line:id_mu_prime}) contains $\Sigma$ in line~\ref{line:id_Sigma}, where $\sigma$ is the entry-wise step function
\begin{equation}\label{eq:step_function}
    \sigma(v) \coloneq \begin{cases}
        1, & \text{if} \quad v > 0,\\
        0, & \text{if} \quad v \leq 0.
    \end{cases}
\end{equation}

\textit{Step~\ref{step:impl_func_thm} (Implicit differentiation of the dual solution).} By~\eqref{eq:dual_to_primal}, the Jacobians of $p(\mu, \theta)$ at $(\mu^*, \theta)$ are $-P^* G^\top$ with respect to $\mu$ and $P^* \Lambda$ with respect to $\theta$. Hence, differentiating~\eqref{eq:fixed_point} via the chain rule~\cite[Lemma~5]{Bolte2021-cs} yields the elements $J_{\mathcal{F},\mu} \in \mathcal{J}_\mu \mathcal{F}$, $J_{\mathcal{F},x} \in \mathcal{J}_x \mathcal{F}$, and $J_{\mathcal{F},\theta} \in \mathcal{J}_\theta \mathcal{F}$ at $(\mu^*, x, \theta)$ in lines~\ref{line:id_U}--\ref{line:id_V}, where $E \coloneq (0_{N \times N}, I_N)$ is the Jacobian of $b(x)$ scaled by $\Delta t$. Since $G$ has full row rank and $P^* \succ 0$, $J_{\mathcal{F},\mu}$ is invertible by an argument analogous to~\cite{Zuliani2023-po}. Hence, by the IFT~\cite[Thm.~5]{Bolte2021-ml}, $\mu^*$ is path-differentiable, with the conservative Jacobians in lines~\ref{line:id_Jmu}--\ref{line:id_Jmu_theta}. Note that $J_{\mathcal{F},\mu}$ has size $2N$, rather than $N|\mathcal{C}|$ as in the primal route.

\textit{Step~\ref{step:primal_sol_jacs} (Conservative Jacobians of the primal solution).} Since $p^*$ is an explicit function of $\mu^*$ and $\theta$ by~\eqref{eq:dual_to_primal}, the chain rule maps the dual Jacobians to $J_{p^*,x} \in \mathcal{J}_x p^*(x,\theta)$ and $J_{p^*,\theta} \in \mathcal{J}_\theta p^*(x,\theta)$ in lines~\ref{line:id_Jpx}--\ref{line:id_Jptheta}.

\ifarxiv
\newpage
\subsection{Proof of Theorem~\ref{thm:conv_to_crit_point}}\label{sec:apx_guarantee_proof}

The proof adapts the analysis of~\cite{Zuliani2023-po}, developed for nested optimization problems with quadratic lower levels, to our nested v-GNEs, whose equivalent social-welfare problems have entropic, hence non-quadratic, costs. We first recall tools from nonsmooth analysis, then establish the regularity of the v-GNE map $p^*$ through the dual characterization of Algorithm~\ref{alg:impl_diff}, and finally prove Theorem~\ref{thm:conv_to_crit_point}.

\paragraph{Preliminaries}

\begin{lemma}[{\cite[Prop.~2]{Bolte2021-cs}}]\label{lem:definable_path_diff}
    Locally Lipschitz functions definable in an o-minimal structure are path-differentiable.
\end{lemma}

\begin{lemma}[{Chain rule~\cite[Lemma~5]{Bolte2021-cs}}]\label{lem:chain_rule}
    Let $\psi_1: \mathbb{R}^{n} \to \mathbb{R}^{m}$ and $\psi_2: \mathbb{R}^{m} \to \mathbb{R}^{l}$ be path-differentiable with conservative Jacobians $\mathcal{J}\psi_1$ and $\mathcal{J}\psi_2$. Then $\psi_2 \circ \psi_1$ is path-differentiable, and $\{AB \mid A \in \mathcal{J}\psi_2(\psi_1(z)),\, B \in \mathcal{J}\psi_1(z)\}$ is a conservative Jacobian for $\psi_2 \circ \psi_1$ at $z \in \mathbb{R}^n$.
\end{lemma}

Since sums, products, and stackings of path-differentiable maps are path-differentiable, with the corresponding rules for their conservative Jacobians~\cite{Bolte2021-cs}, we also apply Lemma~\ref{lem:chain_rule} to maps of several arguments. Moreover, path-differentiability is a local property, so all these statements also hold for maps defined on open sets, such as $\mathcal{O}$ and $\mathcal{U}$ below.

\begin{theorem}[{Implicit function theorem~\cite[Thm.~5]{Bolte2021-ml}}]\label{thm:implicit_function}
    Let $f: \mathbb{R}^{m} \times \mathbb{R}^{n} \to \mathbb{R}^{m}$ be locally Lipschitz and definable, with convex-valued conservative Jacobian $\mathcal{J}f$, and let $f(\tilde{z}, \tilde{\eta}) = 0$. Assume that $U$ is invertible for every $[U \;\, V] \in \mathcal{J}f(\tilde{z}, \tilde{\eta})$, with $U \in \mathbb{R}^{m \times m}$. Then there exist open neighborhoods $\mathcal{U}$ of $\tilde{z}$ and $\mathcal{V}$ of $\tilde{\eta}$, and a definable, path-differentiable map $z^*: \mathcal{V} \to \mathcal{U}$, such that $f(z^*(\eta), \eta) = 0$ for all $\eta \in \mathcal{V}$, and
    \begin{equation}\label{eq:IFT_jacobian}
        \mathcal{J}z^*(\eta) \coloneq \big\{ -U^{-1} V \;\big|\; [U \;\, V] \in \mathcal{J}f(z^*(\eta), \eta) \big\}
    \end{equation}
    is a conservative Jacobian for $z^*$.
\end{theorem}

\begin{theorem}[{\cite[Thm.~6.2]{davis2018stochastic}, \cite{Bolte2021-cs}}]\label{thm:conv_to_crit_point_general}
    Let $f: \mathbb{R}^n \to \mathbb{R}$ be locally Lipschitz and definable, with a definable conservative Jacobian $\mathcal{J}f$, and let $\mathcal{K} \subset \mathbb{R}^n$ be a nonempty polytope. Consider the projected conservative-gradient iteration
    \begin{equation}\label{eq:proj_grad_general}
        z^{(k+1)} = \Pi_\mathcal{K}\big(z^{(k)} - \alpha^{(k)} J^{(k)}\big), \quad J^{(k)} \in \mathcal{J}f(z^{(k)}),
    \end{equation}
    with $z^{(0)} \in \mathcal{K}$ and step sizes $\{\alpha^{(k)}\}_{k \in \mathbb{N}}$ satisfying the Robbins--Monro conditions in Theorem~\ref{thm:conv_to_crit_point}. If $\sup_{k} \|z^{(k)}\| < \infty$, then every accumulation point $\tilde{z}$ of $\{z^{(k)}\}_{k \in \mathbb{N}}$ is a critical point of $f$ over $\mathcal{K}$, i.e., $0 \in \mathcal{J}f(\tilde{z}) + \mathcal{N}_\mathcal{K}(\tilde{z})$, and the sequence of objective values $\{f(z^{(k)})\}_{k \in \mathbb{N}}$ converges.
\end{theorem}

\paragraph{Regularity of the incentive-response map}

Throughout this part, we drop the time index and use the notation of Algorithm~\ref{alg:impl_diff}, i.e., $G = (G_\rmeq, G_\rmiq)$, $b(x)$, $\mathcal{M}$, $\Pi_\mathcal{M}$, $E$, and $p(\mu, \theta)$ in~\eqref{eq:dual_to_primal}. The Lagrange dual function of the social welfare maximization over $\mathcal{Z}(x)$ is
\begin{equation}\label{eq:dual_function}
    q(\mu, x, \theta) \coloneq \mathbbm{1}^\top p(\mu, \theta) + b(x)^\top \mu,
\end{equation}
whose gradient and Hessian with respect to $\mu$ read
\begin{equation}\label{eq:dual_derivatives}
    \begin{split}
        \nabla_\mu q &= b(x) - G\, p(\mu, \theta),\\
        \nabla^2_{\mu\mu} q &= G \, \rmdiag\big(p(\mu, \theta)\big) \, G^\top.
    \end{split}
\end{equation}

\begin{lemma}[Continuity]\label{lem:p_continuous}
    The v-GNE mapping $p^*$ is continuous and definable on $\mathbb{R}^N_+ \times \mathbb{R}^N$.
\end{lemma}

\begin{proof}
    The proof of Lemma~\ref{lem:exist_unique} does not use $\theta \in [0,1]^N$, so $p^*(x, \theta)$ is the unique maximizer of $W(\cdot, \theta)$ over $\mathcal{Z}(x)$ for all $(x, \theta) \in \mathbb{R}^N_+ \times \mathbb{R}^N$. The map $x \mapsto \mathcal{Z}(x)$ is polyhedral, with right-hand side affine in $x$, and nonempty and compact-valued on $\mathbb{R}^N_+$; hence, it is continuous there~\cite{rockafellar1998variational}. Since $W$ is continuous, Berge's maximum theorem~\cite{rockafellar1998variational} and uniqueness yield continuity of $p^*$. Finally, the graph of $p^*$ is $\{(x, \theta, p) \mid p \in \mathcal{Z}(x),\ W(y, \theta) \leq W(p, \theta)\ \forall y \in \mathcal{Z}(x)\}$, which is definable in $\mathbb{R}_{\exp}$, since $W$ is, with $0 \log 0 \coloneq 0$~\cite{coste1999introduction}.
\end{proof}

\begin{lemma}[LICQ]\label{lem:LICQ}
    If $d^\rmpu \neq 0$, then $G$ has full row rank, i.e., the constraints of $\mathcal{Z}(x)$ satisfy the LICQ for any $x \in \mathbb{R}^N$.
\end{lemma}

\begin{proof}
    Let $a, c \in \mathbb{R}^N$ satisfy $G_\rmeq^\top a + G_\rmiq^\top c = 0$. The entry of this vector associated with $p^{j,\emptyset}$ is $a^j$, since $G_\rmiq = \Xi D$ does not involve the leaving options; hence, $a = 0$. The entry associated with $p^{j,i}$, $i \in \mathcal{R}$, is then $d^{\rmpu,j} c^i$. Choosing $j$ with $d^{\rmpu,j} > 0$ yields $c = 0$, so $G$ has full row rank.
\end{proof}

\begin{lemma}[Dual solution]\label{lem:dual_solution}
    Let $x \in \mathbb{R}^N_{>0}$, $\theta \in \mathbb{R}^N$, and $d^\rmpu \neq 0$. Then $q(\cdot, x, \theta)$ admits a unique minimizer $\mu^*(x, \theta)$ over $\mathcal{M}$, which is the unique solution of $\mathcal{F}(\mu, x, \theta) = 0$, and $p^*(x, \theta) = p(\mu^*(x, \theta), \theta)$, with $p$ in~\eqref{eq:dual_to_primal}.
\end{lemma}

\begin{proof}
    Let $\hat{p}$ be given by $\hat{p}^{j,i} = \varepsilon$ for $i \in \mathcal{R}$ and $\hat{p}^{j,\emptyset} = 1 - N\varepsilon$, for all $j \in \mathcal{R}$. Since $x > 0$, for $\varepsilon > 0$ small enough, $\hat{p} > 0$ and $G_\rmiq \hat{p} = \varepsilon (\mathbbm{1}^\top d^\rmpu) \mathbbm{1} < x/\Delta t$. Hence, Slater's condition holds for the (concave, affinely constrained) social welfare maximization, so strong duality holds and the dual minimum over $\mathcal{M}$ is attained. Stationarity of the Lagrangian yields $p^* = p(\mu^*, \theta)$ as in~\eqref{eq:dual_to_primal}. By~\eqref{eq:dual_to_primal}, $p(\mu, \theta) > 0$ for all $\mu$; with Lemma~\ref{lem:LICQ}, \eqref{eq:dual_derivatives} gives $\nabla^2_{\mu\mu} q \succ 0$, so $q(\cdot, x, \theta)$ is strictly convex and its minimizer is unique. Finally, $\mu$ minimizes the convex differentiable $q(\cdot, x, \theta)$ over the closed convex set $\mathcal{M}$ if and only if $\mu = \Pi_\mathcal{M}(\mu - \nabla_\mu q(\mu, x, \theta))$~\cite{Beck2017-pe}, which, by~\eqref{eq:fixed_point} and~\eqref{eq:dual_derivatives}, is equivalent to $\mathcal{F}(\mu, x, \theta) = 0$.
\end{proof}

\begin{lemma}\label{lem:F_definable}
    The map $\mathcal{F}$ in~\eqref{eq:fixed_point} is locally Lipschitz and definable in $\mathbb{R}_{\exp}$, hence path-differentiable. Moreover, with $\mu' \coloneq \mu + G p(\mu, \theta) - b(x)$ and $P \coloneq \rmdiag(p(\mu, \theta))$, a conservative Jacobian of $\mathcal{F}$ at $(\mu, x, \theta)$ is
    \begin{equation}\label{eq:cons_jac_F}
        \mathcal{J}\mathcal{F} \coloneq \big\{ [J_{\mathcal{F},\mu} \;\, J_{\mathcal{F},x} \;\, J_{\mathcal{F},\theta}] \;\big|\; \Sigma \in \mathcal{J}\Pi_\mathcal{M}(\mu') \big\},
    \end{equation}
    where $J_{\mathcal{F},\mu}$, $J_{\mathcal{F},x}$, $J_{\mathcal{F},\theta}$ are as in lines~\ref{line:id_U}--\ref{line:id_V} of Algorithm~\ref{alg:impl_diff}, with $P^*$ replaced by $P$, and $\mathcal{J}\Pi_\mathcal{M}(\mu') \coloneq \{ \rmdiag(\mathbbm{1}_N, s) \mid s \in [0,1]^N,\ s^i = \sigma(\mu'^{\,i}_\rmiq) \text{ if } \mu'^{\,i}_\rmiq \neq 0 \}$. In particular, $\Sigma$ in line~\ref{line:id_Sigma} belongs to $\mathcal{J}\Pi_\mathcal{M}(\mu')$, and $\mathcal{J}\mathcal{F}$ is definable.
\end{lemma}

\begin{proof}
    The map $\Pi_\mathcal{M}(\mu) = (\mu_\rmeq, \max(0, \mu_\rmiq))$ is piecewise linear, hence semialgebraic and definable in $\mathbb{R}_{\exp}$~\cite{coste1999introduction}; the entry-wise exponential is definable in $\mathbb{R}_{\exp}$, which is o-minimal by~\cite{wilkie1996model}; and $r(\theta) = \Lambda \theta - w$ and $b(x)$ are affine. Since compositions and sums of definable maps are definable~\cite{coste1999introduction}, and compositions of locally Lipschitz maps are locally Lipschitz, $\mathcal{F}$ is locally Lipschitz and definable, hence path-differentiable by Lemma~\ref{lem:definable_path_diff}. Since $\Pi_\mathcal{M}$ acts entry-wise, $\mathcal{J}\Pi_\mathcal{M}$, the product of the Clarke subdifferentials of its entries, is a conservative Jacobian for it~\cite{Bolte2021-cs}, and is semialgebraic, hence definable. The Jacobians of the smooth map $(\mu, x, \theta) \mapsto \mu'$ are $I - G P G^\top$, $-E/\Delta t$, and $G P \Lambda$, by~\eqref{eq:dual_to_primal}. Lemma~\ref{lem:chain_rule} yields~\eqref{eq:cons_jac_F}, definable as a composition of definable maps, and~\eqref{eq:step_function} gives the claim on~$\Sigma$.
\end{proof}

\begin{lemma}[Invertibility of $J_{\mathcal{F},\mu}$]\label{lem:inv_U}
    Let $d^\rmpu \neq 0$. For any $(\mu, x, \theta)$, every $J_{\mathcal{F},\mu}$ such that $[J_{\mathcal{F},\mu} \;\, J_{\mathcal{F},x} \;\, J_{\mathcal{F},\theta}] \in \mathcal{J}\mathcal{F}(\mu, x, \theta)$ in~\eqref{eq:cons_jac_F} is invertible.
\end{lemma}

\begin{proof}
    Let $J_{\mathcal{F},\mu} = \Sigma(I - H) - I$, with $\Sigma = \rmdiag(s)$, $s \in [0,1]^{2N}$, and $H \coloneq G P G^\top$, and let $J_{\mathcal{F},\mu} v = 0$, i.e., $\Sigma(I - H)v = v$. Let $\mathcal{A} \coloneq \{k \mid s^k > 0\}$, which contains all indices of $\mu_\rmeq$. For $k \notin \mathcal{A}$, the row associated with $k$ gives $v^k = 0$. For $k \in \mathcal{A}$, it gives $(Hv)^k = -\tfrac{1 - s^k}{s^k} v^k$. Since $v$ is supported on $\mathcal{A}$, these rows read $\big(G^\mathcal{A} P (G^\mathcal{A})^\top + \Gamma\big) v^\mathcal{A} = 0$, where $\Gamma \coloneq \rmdiag\big(\tfrac{1 - s^k}{s^k}\big)_{k \in \mathcal{A}} \succeq 0$ and $G^\mathcal{A}$ collects the rows of $G$ indexed by $\mathcal{A}$. By Lemma~\ref{lem:LICQ}, $G^\mathcal{A}$ has full row rank, and $P \succ 0$ by~\eqref{eq:dual_to_primal}. Hence, $G^\mathcal{A} P (G^\mathcal{A})^\top + \Gamma \succ 0$, so $v = 0$.
\end{proof}

\noindent Lemma~\ref{lem:inv_U} covers the active ($s^k = 1$), inactive ($s^k = 0$), and weakly active ($s^k \in [0,1]$) constraints of~\cite{Zuliani2023-po} at once, which is required since Theorem~\ref{thm:implicit_function} asks for invertibility of every element of the conservative Jacobian.

\begin{lemma}\label{lem:p_path_diff}
    Let $d^\rmpu \neq 0$. Then $\mu^*$ and $p^*$ are definable and path-differentiable on $\mathcal{O} \coloneq \mathbb{R}^N_{>0} \times \mathbb{R}^N$, and $J_{\mu^*,x}, J_{\mu^*,\theta}$ and $J_{p^*,x}, J_{p^*,\theta}$ in lines~\ref{line:id_Jmu}--\ref{line:id_Jptheta} of Algorithm~\ref{alg:impl_diff} are elements of definable conservative Jacobians of $\mu^*$ and $p^*$ at $(x, \theta)$.
\end{lemma}

\begin{proof}
    By Lemma~\ref{lem:dual_solution}, the graph of $\mu^*$ on $\mathcal{O}$ is $\{(x, \theta, \mu) \mid (x, \theta) \in \mathcal{O},\ \mathcal{F}(\mu, x, \theta) = 0\}$, which is definable by Lemma~\ref{lem:F_definable}; hence, $\mu^*$ is definable on $\mathcal{O}$. Fix $(\tilde{x}, \tilde{\theta}) \in \mathcal{O}$. By Lemma~\ref{lem:dual_solution}, $\mathcal{F}(\mu^*(\tilde{x}, \tilde{\theta}), \tilde{x}, \tilde{\theta}) = 0$. The conservative Jacobian~\eqref{eq:cons_jac_F} is convex-valued, since $\mathcal{J}\Pi_\mathcal{M}(\mu')$ is a product of intervals and~\eqref{eq:cons_jac_F} is affine in $\Sigma$. Hence, by Lemmas~\ref{lem:F_definable} and~\ref{lem:inv_U}, Theorem~\ref{thm:implicit_function} applies with $z = \mu$ and $\eta = (x, \theta)$, and yields a path-differentiable $z^*$ with $\mathcal{F}(z^*(x, \theta), x, \theta) = 0$ on a neighborhood of $(\tilde{x}, \tilde{\theta})$, which we take inside the open set $\mathcal{O}$. By the uniqueness in Lemma~\ref{lem:dual_solution}, $z^* = \mu^*$ on this neighborhood. By locality, $\mu^*$ is path-differentiable on $\mathcal{O}$, and, by~\eqref{eq:IFT_jacobian} and~\eqref{eq:cons_jac_F}, lines~\ref{line:id_Jmu}--\ref{line:id_Jmu_theta} return an element of its conservative Jacobian, which is definable as the image of the definable set~\eqref{eq:cons_jac_F} under a definable map. Finally, $p^* = p(\mu^*, \theta)$ by Lemma~\ref{lem:dual_solution}, with $p(\cdot, \cdot)$ in~\eqref{eq:dual_to_primal} smooth and definable, with Jacobians $-P^* G^\top$ and $P^* \Lambda$; hence, $p^*$ is definable, and Lemma~\ref{lem:chain_rule} yields lines~\ref{line:id_Jpx}--\ref{line:id_Jptheta}.
\end{proof}

\paragraph{Regularity of the cost-to-go}

By Lemma~\ref{lem:p_continuous}, the state update map $\mathcal{T}$ in~\eqref{eq:state_update_map} is continuous and definable on $\mathbb{R}^N_+ \times \mathbb{R}^N$, and maps it into $\mathbb{R}^N_+$, since $p^*(x_t, \theta_t) \in \mathcal{Z}(x_t)$ gives $\Delta t\, \Xi D p^*(x_t, \theta_t) \leq x_t$. Hence, for a fixed $x_0 \in \mathbb{R}^N_+$, the states $x_t(\boldsymbol{\theta})$ in~\eqref{eq:cost_to_go}--\eqref{eq:incentive_planning_problem_simplified} are well defined, continuous, and definable functions of $\boldsymbol{\theta} \in \mathbb{R}^{TN}$, and so is $m(\boldsymbol{\theta}) \coloneq \min_{t \in \{0, \dots, T\}} \min_{i \in \mathcal{R}} x^i_t(\boldsymbol{\theta})$. Lemma~\ref{lem:p_path_diff}, instead, requires $x \in \mathbb{R}^N_{>0}$: for $x^i = 0$, $p^{*,j,i} = 0$ for all $j$ with $d^{\rmpu,j} > 0$, contradicting $p^* > 0$ in~\eqref{eq:dual_to_primal}, so the dual minimum is not attained. We therefore work on the open set
\begin{equation}\label{eq:pos_set}
    \mathcal{U} \coloneq \{ \boldsymbol{\theta} \in \mathbb{R}^{TN} \mid m(\boldsymbol{\theta}) > 0 \},
\end{equation}
of incentives that keep all states positive. The assumption of Theorem~\ref{thm:conv_to_crit_point} states that all iterates lie in $\mathcal{K}_\delta \coloneq \{\boldsymbol{\theta} \in \Omega \mid m(\boldsymbol{\theta}) \geq \delta\}$, which is compact, as a closed subset of the compact set $\Omega$, and contained in $\mathcal{U}$.

\begin{remark}\label{rem:positive_states}
    If $x_0 > 0$ and $d^\rmdo_t > 0$ for all $t \in \mathbb{T}$, then $x_{t+1} \geq \Delta t\, d^\rmdo_t$ for all $t$ by~\eqref{eq:state_update_map}, so $\mathcal{U} = \mathbb{R}^{TN}$, and the assumption of Theorem~\ref{thm:conv_to_crit_point} holds with $\delta = \min\{\min_i x^i_0,\ \Delta t \min_{t,i} d^{\rmdo,i}_t\}$.
\end{remark}

\begin{lemma}\label{lem:V0_path_diff}
    Let $x_0 \in \mathbb{R}^N_{>0}$. Then $V_0(x_0, \cdot)$ is definable and path-differentiable on $\mathcal{U}$, with a definable conservative Jacobian $\mathcal{J}_{\boldsymbol{\theta}} V_0(x_0, \cdot)$, and, for any $\boldsymbol{\theta} \in \mathcal{U}$, the output $J^{(0)}_{V,\boldsymbol{\theta}}$ of Algorithm~\ref{alg:bwrd_prop} belongs to $\mathcal{J}_{\boldsymbol{\theta}} V_0(x_0, \boldsymbol{\theta})$.
\end{lemma}

\begin{proof}
    At intervals with $d^\rmpu_t \neq 0$, by Lemma~\ref{lem:p_path_diff}, the maps $\mathcal{T}$ in~\eqref{eq:state_update_map} and $g$ in~\eqref{eq:stage_cost_state} are definable and path-differentiable on $\mathcal{O}$, as compositions of $p^*$ with polynomial maps (Lemma~\ref{lem:chain_rule}). At intervals with $d^\rmpu_t = 0$, instead, $D = 0$, so $\mathcal{S} \equiv 0$, and $\mathcal{T}(x, \theta_t) = x + \Delta t\, d^\rmdo_t$ and $g(x, \theta_t) = \gamma_R \|\theta_t\|_2^2$ are smooth; the terms involving $J^{(t)}_{p^*,x}$ and $J^{(t)}_{p^*,\theta}$ in~\eqref{eq:jac_g}--\eqref{eq:jac_T} are multiplied by $D = 0$ and vanish, so Lemma~\ref{lem:p_path_diff} is not needed. In both cases, \eqref{eq:jac_g}--\eqref{eq:jac_T} follow from Lemma~\ref{lem:chain_rule} applied to~\eqref{eq:state_update_map}--\eqref{eq:stage_cost_state}, so $[J^{(t)}_{g,x} \;\, J^{(t)}_{g,\boldsymbol{\theta}}] \in \mathcal{J}g(x_t, \theta_t)$ and $[J^{(t)}_{\mathcal{T},x} \;\, J^{(t)}_{\mathcal{T},\boldsymbol{\theta}}] \in \mathcal{J}\mathcal{T}(x_t, \theta_t)$, viewed as functions of $(x_t, \boldsymbol{\theta})$, with definable conservative Jacobians.

    Define $\mathcal{U}_T \coloneq \mathbb{R}^N_{>0} \times \mathbb{R}^{TN}$ and, for $t = T-1, \dots, 0$, $\mathcal{U}_t \coloneq \{(x, \boldsymbol{\theta}) \in \mathbb{R}^N_{>0} \times \mathbb{R}^{TN} \mid (\mathcal{T}(x, \theta_t), \boldsymbol{\theta}) \in \mathcal{U}_{t+1}\}$, which are open by continuity of $\mathcal{T}$. We show by backward induction that, for $t = T, \dots, 0$, $V_t$ is definable and path-differentiable on $\mathcal{U}_t$, with a definable conservative Jacobian containing $[J^{(t)}_{V,x} \;\, J^{(t)}_{V,\boldsymbol{\theta}}]$ at $(x_t, \boldsymbol{\theta})$. For $t = T$, $V_T = h$ is smooth with gradient as in line~\ref{line:bp_init_x} of Algorithm~\ref{alg:bwrd_prop}. For $t < T$, \eqref{eq:cost_to_go} writes $V_t$ on $\mathcal{U}_t$ as the sum of $g$ and the composition of $V_{t+1}$ with $(x, \boldsymbol{\theta}) \mapsto (\mathcal{T}(x, \theta_t), \boldsymbol{\theta})$, which maps $\mathcal{U}_t$ into $\mathcal{U}_{t+1}$; all these maps are definable and path-differentiable, and Lemma~\ref{lem:chain_rule} yields exactly lines~\ref{line:bp_Vx}--\ref{line:bp_Vtheta}. Finally, $\boldsymbol{\theta} \in \mathcal{U}$ if and only if $(x_0, \boldsymbol{\theta}) \in \mathcal{U}_0$, so, with $x_0$ fixed, $V_0(x_0, \cdot)$ is definable and path-differentiable on $\mathcal{U}$, and $J^{(0)}_{V,\boldsymbol{\theta}} \in \mathcal{J}_{\boldsymbol{\theta}} V_0(x_0, \boldsymbol{\theta})$.
\end{proof}

\paragraph{Proof of Theorem~\ref{thm:conv_to_crit_point}}

Since $V_0(x_0, \cdot)$ is only defined as a path-differentiable function on $\mathcal{U}$, we apply Theorem~\ref{thm:conv_to_crit_point_general} to a function that coincides with it near $\mathcal{K}_\delta$. Note that $x_0 \geq \delta \mathbbm{1} > 0$ by assumption, so Lemma~\ref{lem:V0_path_diff} applies. Let $\psi(s) \coloneq \min\{1, \max\{0,\ 4s/\delta - 2\}\}$, which is piecewise linear, equal to $0$ for $s \leq \delta/2$ and to $1$ for $s \geq 3\delta/4$, and define
\begin{equation}\label{eq:cutoff}
    f(\boldsymbol{\theta}) \coloneq \begin{cases}
        \psi\big(m(\boldsymbol{\theta})\big)\, V_0(x_0, \boldsymbol{\theta}), & \text{if } \boldsymbol{\theta} \in \mathcal{U},\\
        0, & \text{otherwise}.
    \end{cases}
\end{equation}
The support of $\psi \circ m$ is contained in the closed set $\{m \geq \delta/2\} \subset \mathcal{U}$, so $f$ vanishes on a neighborhood of every point outside it. On $\mathcal{U}$, $m$ is locally Lipschitz and definable, as a minimum of compositions of the locally Lipschitz definable maps $\mathcal{T}$. By Lemma~\ref{lem:V0_path_diff}, $f$ is locally Lipschitz and definable on $\mathbb{R}^{TN}$, and admits the definable conservative Jacobian $\mathcal{J}f$ given by the product rule~\cite{Bolte2021-cs} on $\mathcal{U}$ and by $\{0\}$ elsewhere. On the open set $\{m > 3\delta/4\}$, which contains $\mathcal{K}_\delta$, $\psi \circ m \equiv 1$, so $f = V_0(x_0, \cdot)$ and $\mathcal{J}f = \mathcal{J}_{\boldsymbol{\theta}} V_0(x_0, \cdot)$ there.

We now apply Theorem~\ref{thm:conv_to_crit_point_general} with this $f$, $\mathcal{K} = \Omega$, and $z^{(k)} = \boldsymbol{\theta}^{(k)}$. First, $f$ is locally Lipschitz and definable with a definable conservative Jacobian. Second, $\Omega = [0, \bar{\theta}]^{TN}$ is a nonempty box, hence a polytope. Third, since $\boldsymbol{\theta}^{(k)} \in \mathcal{K}_\delta$ for all $k$ by assumption, Lemma~\ref{lem:V0_path_diff} gives $J^{(k)}_{V,\boldsymbol{\theta}} \in \mathcal{J}_{\boldsymbol{\theta}} V_0(x_0, \boldsymbol{\theta}^{(k)}) = \mathcal{J}f(\boldsymbol{\theta}^{(k)})$ in line~\ref{line:pgd_jacobian} of Algorithm~\ref{alg:proj_grad_desc}, so line~\ref{line:pgd_step} is an instance of~\eqref{eq:proj_grad_general} (with $\varepsilon = 0$, i.e., without termination). Fourth, the step sizes satisfy the Robbins--Monro conditions by assumption. Finally, $\sup_k \|\boldsymbol{\theta}^{(k)}\| < \infty$, as $\boldsymbol{\theta}^{(k)} \in \Omega$ and $\Omega$ is compact. Hence, by Theorem~\ref{thm:conv_to_crit_point_general}, every accumulation point $\tilde{\boldsymbol{\theta}}$ of $\{\boldsymbol{\theta}^{(k)}\}_{k \in \mathbb{N}}$ satisfies $0 \in \mathcal{J}f(\tilde{\boldsymbol{\theta}}) + \mathcal{N}_\Omega(\tilde{\boldsymbol{\theta}})$, and $\{f(\boldsymbol{\theta}^{(k)})\}_{k \in \mathbb{N}}$ converges. Since $\mathcal{K}_\delta$ is closed, $\tilde{\boldsymbol{\theta}} \in \mathcal{K}_\delta$, where $\mathcal{J}f = \mathcal{J}_{\boldsymbol{\theta}} V_0(x_0, \cdot)$ and $f = V_0(x_0, \cdot)$. Therefore, $0 \in \mathcal{J}_{\boldsymbol{\theta}} V_0(x_0, \tilde{\boldsymbol{\theta}}) + \mathcal{N}_\Omega(\tilde{\boldsymbol{\theta}})$, i.e., $\tilde{\boldsymbol{\theta}}$ is a critical point of~\eqref{eq:incentive_planning_problem_simplified}, and $\{V_0(x_0, \boldsymbol{\theta}^{(k)})\}_{k \in \mathbb{N}}$ converges. \hfill $\blacksquare$
\else
\newpage
\fi
\balance
\bibliographystyle{IEEEtran} %max 0.75 pages
\bibliography{bibTeX/IEEEabrv,mybib.bib}

\end{document}